\documentclass[useAMS,usenatbib]{mn2e} 
\pdfoutput=1 
\usepackage{journals}
\usepackage[pdftex]{graphicx,color} 
\usepackage[latin1]{inputenc}
\usepackage{graphics} 
\usepackage{amsfonts} 
\usepackage{amsmath}
\usepackage{multicol} 
\usepackage{layout} 
\usepackage{amssymb}
\usepackage[a4paper,colorlinks=true,pdfstartview=FitV,
linkcolor=red,citecolor=blue,urlcolor=magenta]{hyperref}
\title[MOKA: Strong Lensing in Galaxy Clusters]
{MOKA: a new tool for Strong Lensing Studies}
\author[Giocoli et al. 2011]
{\parbox{\textwidth}{Carlo Giocoli$^{1,2,3}$\thanks{E-mail:
 \href{mailto:cgiocoli@oabo.inaf.it} {cgiocoli@oabo.inaf.it}},
 Massimo Meneghetti$^{1,3}$, Matthias Bartelmann$^{2}$, Lauro Moscardini$^{1,3,4}$,
 Michele Boldrin$^{4}$} \\ \\ 
 $^{1}$ INAF - Osservatorio Astronomico di Bologna, via Ranzani 1, 40127, Bologna,
 Italy \\ 
 $^{2}$ Zentrum f\"ur Astronomie, ITA, Universit\"at Heidelberg, 
 Albert-Ueberle-Str. 2, 69120 Heidelberg, Germany \\ 
 $^{3}$ INFN - Sezione di Bologna, viale Berti Pichat 6/2, 40127,
 Bologna, Italy \\
 $^{3}$ Dipartimento di Astronomia, Universit\`a di Bologna,
 via Ranzani 1, 40127, Bologna, Italy 
 \\ }
\begin{document}
\date{}
\maketitle
\label{firstpage}
\pagerange{\pageref{firstpage}--\pageref{lastpage}} \pubyear{2011}
\begin{abstract}

  Strong  gravitational lensing  is a  powerful tool  for  probing the
  matter  distribution in  the cores  of massive  dark  matter haloes.
  Recent and ongoing analyses of galaxy cluster surveys (MACS, CFHTLS,
  SDSS, SGAS, CLASH, LoCuSS) have  adressed the question of the nature
  of the dark matter  distribution in clusters.  N-body simulations of
  cold  dark-matter haloes  consistently  find that  haloes should  be
  characterized  by  a   concentration-mass  relation  that  decreases
  monotonically with  halo mass,  and populated by  a large  amount of
  substructures, representing the  cores of accreted progenitor halos.
  It is important  for our understanding of dark  matter to test these
  predictions.   We   present  \texttt{MOKA},  a   new  algorithm  for
  simulating  the  gravitational  lensing  signal  from  cluster-sized
  haloes.   It  implements  the  most recent  results  from  numerical
  simulations to create realistic cluster-scale lenses with properties
  independent of numerical  resolution.  We perform systematic studies
  of  the  strong  lensing  cross   section  as  a  function  of  halo
  structures.  We  find that the strong lensing  cross sections depend
  most strongly  on the  concentration and on  the inner slope  of the
  density profile of  a halo, followed in order  of importance by halo
  triaxiality and the presence of a bright central galaxy.

\end{abstract}

\section{Introduction}
Galaxy   clusters    are   an   important    probe   for   dark-matter
properties. According to the standard scenario of structure formation,
galaxy  clusters  as  a   population  are  still  in  their  formation
process. Since  gas cooling cannot substantially  compress dark matter
haloes, their density profiles are dominated by dark matter.

Recent analyses  of strong  and weak lensing  by galaxy  clusters have
found good  consistence with an NFW-like profile  whose projected mass
density  distribution  continuously flattens  towards  the centre,  as
expected from CDM dominated haloes.  However, for an increasing number
of  clusters  \citet{oguri05,oguri09,umetsu11}  have  found  that  the
mass-concentration  relation  seems  to  lie substantially  above  the
relation predicted by CDM  simulations.  Strong lensing by clusters is
sensitive to their internal  structure -- mass distribution within the
Einstein radius\footnote{For  an axially symmetric lens  is defined as
  the radius of a circle  enclosing a mean convergence of $1$.}, which
includes  (i)  presence  of  substructure,  (ii)  asymmetry  in  their
gravitational potential  well, (iii)  ellipticity, (iv) presence  of a
massive and bright  central galaxy (v) inner slope  of the dark matter
density profile.

Apart from weak and strong  lensing, the halo density profile can also
be  constrained from the  velocity dispersion  profile of  the central
galaxy  \citep{sand04,newman09}  and   X-ray  emission  from  the  hot
intra-cluster gas.  While gravitational lensing  does not rely  on any
equilibrium or  dynamical assumptions, methods based on  galaxy or gas
dynamics do, potentially biasing mass and concentration estimates.

In  this paper,  we shall  quantify the  importance of  the structural
parameters of dark matter haloes  for strong lensing signal, using our
new   and  fast   algorithm  \texttt{MOKA}   (\texttt{M}atter  density
distributi\texttt{O}n \texttt{K}ode for gravitation\texttt{A}l lenses)
to create realistic maps of substructured triaxial dark matter haloes,
which are  in perfect agreement  with the measurements done  on galaxy
clusters extracted from numerical simulations. 

In Section \ref{sechprop} we present  the halo properties on which our
algorithm relies. The halo lensing properties are presented in Section
\ref{seclens}  and the strong  lensing signal  dependence on  the halo
modelling is  discussed in Sections  \ref{secresults1}. Discussion and
conclusion are presented in Section \ref{disccon}.

We   adopt  a  $\mathrm{\Lambda   CDM}$  model   with  $\Omega_m=0.3$,
$\Omega_{\Lambda}=0.7$, $h=0.7$  and $\sigma_8=0.9$, consistently with
the Mare Nostrum Universe that we compare some of our results to.

\section{Construction of Realistic Lenses}
\label{sechprop}
Strong gravitational  lensing depends on the  matter distribution near
halo  centres  where  limited  numerical  resolution  may  affect  the
particle   distribution.    Here,   we   present  a   new   algorithm,
\texttt{MOKA},  which   analytically  creates  surface   mass  density
distributions  of  triaxial and  substructured  haloes independent  of
numerical  resolution.   \texttt{MOKA} is  publicly  available at  the
following                                                           url
\href{http://cgiocoli.wordpress.com/research-interests/moka}http://cgiocoli.wordpress.com/research-interests/moka.
The idea  behind this new  algorithm is to construct  realistic lenses
starting  from  a set  of  ingredients taken  from  state  of the  art
numerical simulations.  It not only  accounts for the smooth dark halo
and  stellar  components,  as  earlier studies  do  \citep{vandeven09,
  mandelbaum09}, but also the presence of substructures perturbing the
regular  matter  distribution.   The  procedure is  discussed  in  the
following subsections.

\subsection{Host Halo Contents}
Halo   properties   are  described   by   the   extended  halo   model
\citet{giocoli10b}, developed for the reconstruction of the non-linear
dark matter power spectrum. We briefly list its ingredients here.

\subsubsection{Properties of the dark matter component}
The virial mass of a halo is defined as
\begin{equation}
  M_{vir}         =          \frac{4         \pi}{3}         R_{vir}^3
  \frac{\Delta_{vir}}{\Omega_{m}(z)} \Omega_0 \rho_c\,,
\label{massdef}
\end{equation} 
where  $\rho_c$  represents  the  critical density  of  the  Universe,
$\Omega_0=\Omega_m(0)$  the matter  density parameter  at  the present
time     and    $\Delta_{vir}$     is    the     virial    overdensity
\citep{eke96,bryan98}, $R_{vir}$  symbolizes the virial  radius of the
halo which  defines the dinstance  from the halo centre  that encloses
the desired density contrast.  In  what follows, we summarize the most
recent numerical results built into \texttt{MOKA}.

\begin{itemize}

\item As said above, the  dark matter density distribution in isolated
  haloes is well described by the NFW \citep{navarro96} profile
\begin{equation}
  \rho(r|M_{vir}) = \frac{\rho_s}{r/r_s(1+r/r_s)^2}\,,
\label{eqNFW}
\end{equation}
where $r_s$  is the scale radius, defining  the concentration $c_{vir}
\equiv R_{vir}/r_s$,  and $\rho_s$ is  the dark matter density  at the
scale radius,
\begin{equation}
  \rho_s = \frac{M_{vir}}{4 \pi r_s^3}
  \left[ \ln(1+c_{vir}) - \frac{c_{vir}}{1+c_{vir}}\right]^{-1}\,.
\label{eqrhos}
\end{equation}
Combining the  preceding three equations, we can  explicitly write the
NFW profile as a function of $c_{vir}$ and $M_{vir}$,
\begin{eqnarray}
\rho(r|M_{vir}) &=& \dfrac{c_{vir}^2 R_{vir}^3}{3 r (R_{vir}+c_{vir} r)^2}
\dfrac{\Delta_{vir}}{\Omega_m(z)} \Omega_0 \rho_c \nonumber \\ 
&\times&\left[ \ln(1+c_{vir}) - \frac{c_{vir}}{1+c_{vir}}\right]^{-1}\,.
\end{eqnarray}
We  use this
density profile to model the  dark matter distribution in the clusters
produced by \texttt{MOKA}.

\item The halo concentration is a decreasing function of the host halo
  mass. This  numerical result is  explained in terms  of hierarchical
  clustering    in    CDM    and    the    different    halo-formation
  histories. Already \citet{bond91} and \citet{lacey93}, following the
  extended-\citet{press74} theory,  have found that  the collapse time
  of  dark matter  haloes  depends on  the  halo mass  and that  their
  assembly history  is hierarchical: small systems  collapse at higher
  redshifts than larger  ones \citep{sheth04a,giocoli07a}.  This trend
  is reflected in the mass-concentration relation: at a given redshift
  smaller  haloes are  more concentrated  than larger  ones. Different
  fitting  functions for  numerical mass-concentration  relations have
  been given \citep{bullock01a,neto07,duffy08,gao08}. In this work, we
  use  the  relation  proposed   by  \citet{zhao09}  which  links  the
  concentration of  a given halo  with the time ($t_{0.04}$)  at which
  its main progenitor assembles $4$ percent of its mass,
\begin{equation}
  c_{vir}(M_{vir},z_l) = 4 \left\{ 1 +
  \left[ \frac{t(z_l)}{3.75 t_{0.04}}\right]^{8.4} \right\}^{1/8}\,.
\label{eqzhao}
\end{equation}
The model by \citet{zhao09}  fits well numerical simulations even with
different cosmologies. It seems to be of reasonably general validity.

Due to different assembly histories, haloes with same mass at the same
redshift        may        have        different        concentrations
\citep{navarro96,jing00,wechsler02,zhao03a,zhao03b}.   At  fixed  host
halo mass, the  distribution in concentration is well  fitted by a log
normal distribution function with  a variance $\sigma_{\ln c}$ between
$0.1$ and $0.25$ \citep{jing00,dolag04,sheth04b,neto07}. 

\item Halos are generally not spherical but triaxial due to their tidal interaction with the surrounding density field during their collapse \citep{sheth99b,sheth02}. The correlation of the halo shape with the surrounding environment is expressed as a function of the matter density parameter and of the typical collapse mass. \citet{jing02} have performed a statistical study of halo shapes extracted from numerical simulations, deriving that, if $a$, $b$ and $c$ represent the minor, median and major axes, respectively, empirical relations for $a/c$ and $a/b$ are given by the distribution
\begin{equation}
  p(\lambda) \mathrm{d} \lambda = \frac{1}{\sqrt{2 \pi} \sigma_{\lambda}}
  \exp \left[ - \frac{(\lambda - 0.54 )^2}{2 \sigma_{\lambda}} \right]
  \mathrm{d} \lambda
\end{equation}
for  $\lambda  =  (a/c)  (M_{vir}/M_*(z_l))^{0.07  \Omega(z_l)}$  with
$\sigma_{\lambda}=0.113$ and the  conditional probability for the axis
ratios
\begin{displaymath}
  p\left(a/b| a/c\right) = \left\{ \begin{array}{ll}
    \frac{3}{2(1-r_{min}}) \left[
    1 - \left( \frac{2a/b - 1 - r_{min}}{1-r_{min}} \right)^2 \right]
    \ &a/b \geq r_{min}, \\
    0 &a/b < r_{min}, \\
  \end{array} \right. 
\end{displaymath}
where $r_{min}  = 0.5$ if $a/c<0.5$  else $r_{min} =  a/c$.  As usual,
$M_*(z_l)$ is the non-linear  mass at redshift $z_l$. Simulations with
gas cooling have  produced haloes that tend to  be more spherical than
those in pure DM simulation \citep{kazantzidis04}. The host halo shape
then depends also on the morphology of the stellar mass component.  In
this paper we use the prescriptions by \citet{jing02}, which are based
on  dark matter only  simulations.  However,  \texttt{MOKA} is  a very
flexible  tool, and  the input  distributions  of axis  ratios can  be
easily modified.

\item Haloes  are not smooth, but  characterized by a  large number of
  substructures,   which   may   or   not  host   satellite   galaxies
  \citep{moore99,springel01b,delucia04}. These substructures are cores
  of progenitor  haloes accreted along  the merger tree that  were not
  completely        disrupted         by        tidal        stripping
  \citep{springel01b,gao04,delucia04,vandenbosch05,giocoli08b,shaw07}. Because
  of  different assembly histories  and time  scales for  subhalo mass
  loss,  more  massive  haloes  retain more  substructures  than  less
  massive haloes  at a given  redshift. Likewise, haloes with  a lower
  concentration  (and thus  with a  lower formation  redshift)  are on
  average    more    substructured    than    haloes    with    larger
  concentration. \citet{giocoli10} fitted the subhalo mass function by
  \begin{eqnarray}
  &&\frac{1}{M_{vir}} \frac{\mathrm{d}N(M_{vir},c_{vir},z_l)}{\mathrm{d}m} =
  \nonumber \\
  &&A (1+z_l)^{1/2} \frac{\bar{c}}{c_{vir}}
  m^{\alpha} \exp\left[ - \beta \left( \frac{m}{M_{vir}} \right)^3\right]\,,
\label{eqsubhalomassf}
\end{eqnarray}
where  $A=9.33  \times  10^{-4}$, $\beta=12.2715$,  $\alpha=-0.9$  and
$\bar{c}$ is the  mean concentration of a halo  with mass $M_{vir}$ at
redshift  $z_l$.  We adopt  this  model  because  it incorporates  the
dependences of  the subhalo mass  function on the mass  $M_{vir}$, the
concentration $c_{vir}$  and the redshift  $z_l$ of the host  halo. To
populate  our  haloes  with  substructures  with mass  $m_i$  we  will
randomly  sample  the distribution  down  to  a  minimal subhalo  mass
$m_{min}$.

Subhalo density profiles are modified  by tidal stripping due to close
interactions  with  the  main  halo  smooth  component  and  to  close
encounters  with other  clumps, gravitational  heating,  and dynamical
friction.  Such events  can cause the subhaloes to  lose mass, and may
eventually      result      in      their     complete      disruption
\citep{hayashi03,choi07}.   The  remaining  self-bound subhaloes  will
have density  profiles different from the  NFW shape in  that they are
truncated at the tidal radius \citep{bullock01a}. In order to take the
truncation into account, we model the dark matter density distribution
in   subhaloes    using   truncated   singular    isothermal   spheres
\citep{keeton03},
\begin{displaymath}
  \rho_{sub}(r) = \left\{ \begin{array}{ll}
    \dfrac{\sigma_v^2}{2 \pi G r^2} &r\le R_{sub}, \\
    0  &r>R_{sub} \\
  \end{array} \right.
\end{displaymath}
with  velocity dispersion  $\sigma_v$, and $R_{sub}$ defined as:
\begin{eqnarray}
m_{sub} &=& \int_0^{R_{sub}} 4 \pi r^2 \rho_{sub}(r) \mathrm{d}r \Rightarrow \nonumber \\
R_{sub} &=& \dfrac{G\,m_{sub}}{2\,\sigma_v^2}\,. \label{rsub}
\end{eqnarray}
The velocity  dispersion $\sigma_v$  is related to the
subhalo  temperature by
\begin{equation}
\sigma_v = f(T) \sigma_v^{(R)} + [1-f(T)] \sigma_v^{(T)}(T)\,,
\label{EQsigma}
\end{equation}
where
\begin{equation}
f(T) = \left[ 1 + \left( \dfrac{k T}{\mathrm{keV}}\right)^4 \right]^{-1}\,.
\end{equation}
Based on  the temperature definition in the  spherical collapse model,
we can write
\begin{equation}
  k T = (k T)_{15} M_{15}^{2/3} (1+z) \left( \dfrac{\Delta_{vir} \Omega_0 }{18 \pi^2 \Omega_m(z)} \right)^{1/3}\,
\end{equation}
where $k$ is  Boltzmann's constant, and $M_{15}$ is  the mass in units
of  $10^{15}\,M_{\odot}/h$.  For  temperatures  much  above  $1$  keV,
Eq.~(\ref{EQsigma})  follows  the   observed  relation  between  X-ray
temperature and velocity dispersion by \citet{wu98},
\begin{equation}
\sigma_v^{(T)} = 371.5 \left( \dfrac{k T}{\mathrm{keV}} \right)^{0.56}\,\mathrm{km\,s^{-1}}\,;
\end{equation}
while for $T\ll1$ keV it reduces to the relation
\begin{equation}
\sigma_v^{(R)} = \left( \dfrac{G m_{sub} H(z) \Delta_{vir}^{1/2}}{4} \right)^{1/3}
\end{equation}
that can  be obtained combining  Eqs.~(\ref{rsub}) and (\ref{massdef})
identifying $R_{sub}=R_{vir}$.

Let us define $R_t$ as the radius at which the subhalo mean density is
of  the   order  of  the  mean   density  of  the   main  halo  within
$r$. Following \citet{tormen98b} we can write it as:
\begin{equation}
R_t = r \left[\dfrac{m_{sub}}{(2-\partial \ln M_{vir}(r) / \partial \ln r)M_{vir}(r)} \right]^{1/3}\,,
\end{equation}
where $m_{sub}$  and $r$ represent  the subhalo mass and  its distance
from  the  host halo  centre,  and  $M_{vir}(r)$  the host  halo  mass
profile.    Truncating  subhaloes   at  $R_{sub}$   does   not  create
discontinuities  in  the  convergence  map  because  $R_t(r)  \lesssim
R_{sub} $.  It  preserves the total subhalo mass  fraction in hosts as
found in  numerical simulations \citep{gao04,delucia04,giocoli10}.  If
the  host  halo  matter   density  distribution  is  described  by  an
NFW-profile, the  subhalo tidal radius  as a function of  the distance
from the  host halo centre  can be analytically estimated.   In Figure
\ref{rtidalrsub} we  show the ratio  between the subhalo  tidal radius
and the  radius which encloses the  subhalo mass as a  function of the
distance from  the host halo centre.   From the figure  we notice that
for  $r<0.75  R_{vir}$ $R_t\ll  R_{sub}$  while  when  the subhalo  is
located near at the virial radius the ratio tends to unity.

\begin{figure}
\includegraphics[width=\hsize]{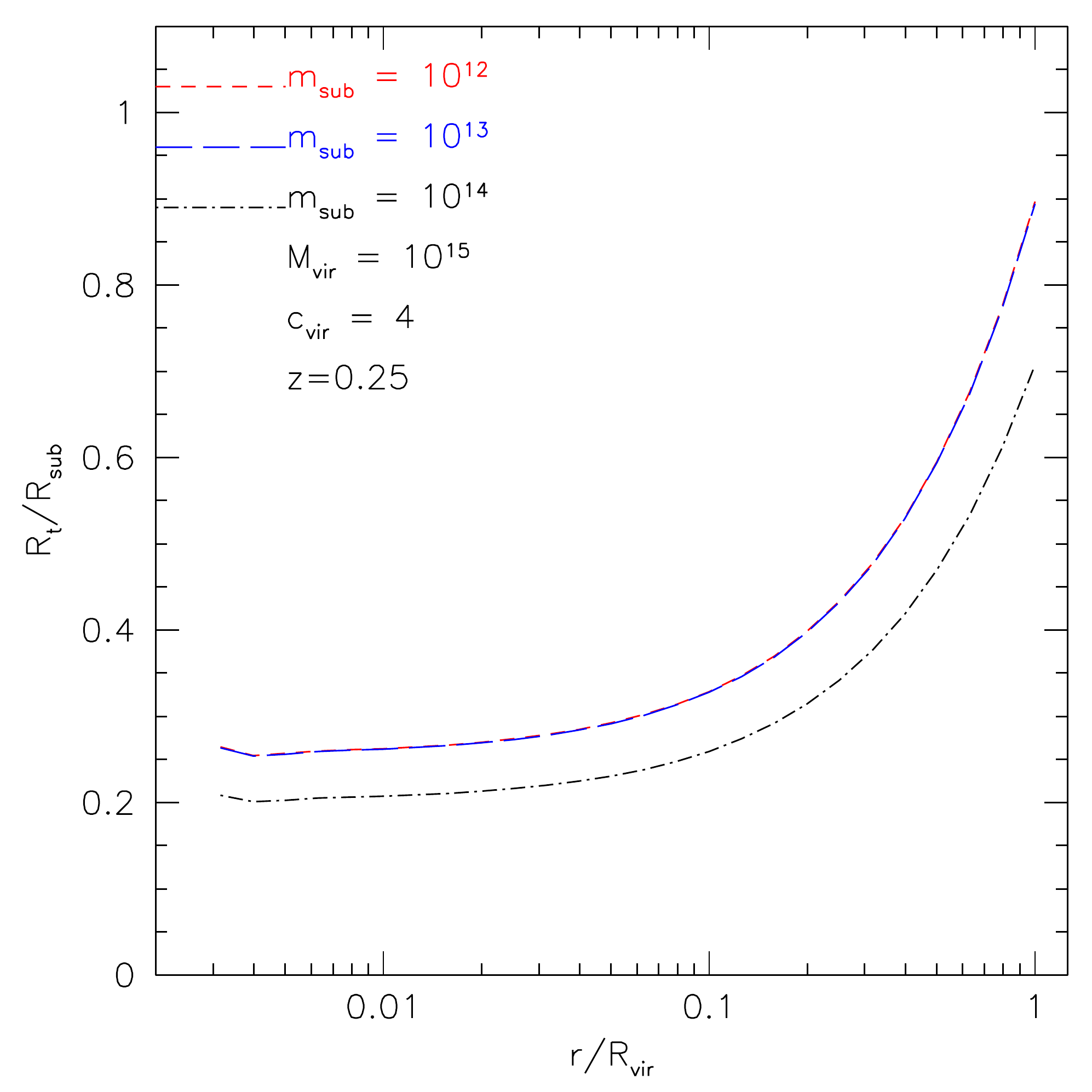}
\caption{\label{rtidalrsub}Ratio between the  subhalo tidal radius and
  the one which  encloses its mass as a function  of the distance from
  the  host halo  centre. In  the  figure we  show the  case of  three
  subhaloes located in a host halo with mass $10^{15}\,M_{\odot}/h$ at
  redshift $z=0.25$ with concentration $c_{vir}=4$.}
\end{figure}

The  SIS profile profile  well represents  galaxy density  profiles on
scales relevant for strong lensing. Previously, different authors have
used this  model to characterize  the lensing signal  by substructures
after  stripping  \citep{metcalf01}.  Nonetheless, additional  subhalo
density  profiles  will  be  implemented  in  \texttt{MOKA}  to  allow
truncating the subhalo profile more smoothly.

\item The spatial distribution of subhaloes tends to follow the smooth
  dark matter distribution of the  host halo. However, clumps near the
  host  halo   center  are  easily  destroyed,  due   to  their  tidal
  interaction  with   the  main   halo.  Thus,  the   spatial  subhalo
  distribution is less  concentrated than the NFW profile  of the host
  halo. \citet{gao04} have studied this using a cosmological numerical
  simulation   and   found  that   the   cumulative  spatial   density
  distribution of clumps in host haloes is well described by
\begin{equation}
\frac{n(<x)}{N_{tot}} = \frac{(1+\alpha' c_{vir}) x^{\beta'}}{(1+\alpha' c_{vir}x^{2})}\,
\label{gaoEQ}
\end{equation}
where  $x$ is the  distance to  the host  halo centre  in unit  of the
virial radius, $N_{tot}$ is the total number of subhaloes in the host,
$\alpha'=0.244$ and $\beta'=2.75$. In  the Figure \ref{gaoFIG} we show
previouse equation for different value of the host halo concentration,
as indicated in  the label.  To place sublumps in  the host, we sample
this  distribution and  randomly assign  position angles  $\theta$ and
$\phi$ on  a sphere.  A  similar equation has  been used to  model the
satellite galaxy density distribution by \citet{vandenbosch04}.

\begin{figure}
\includegraphics[width=\hsize]{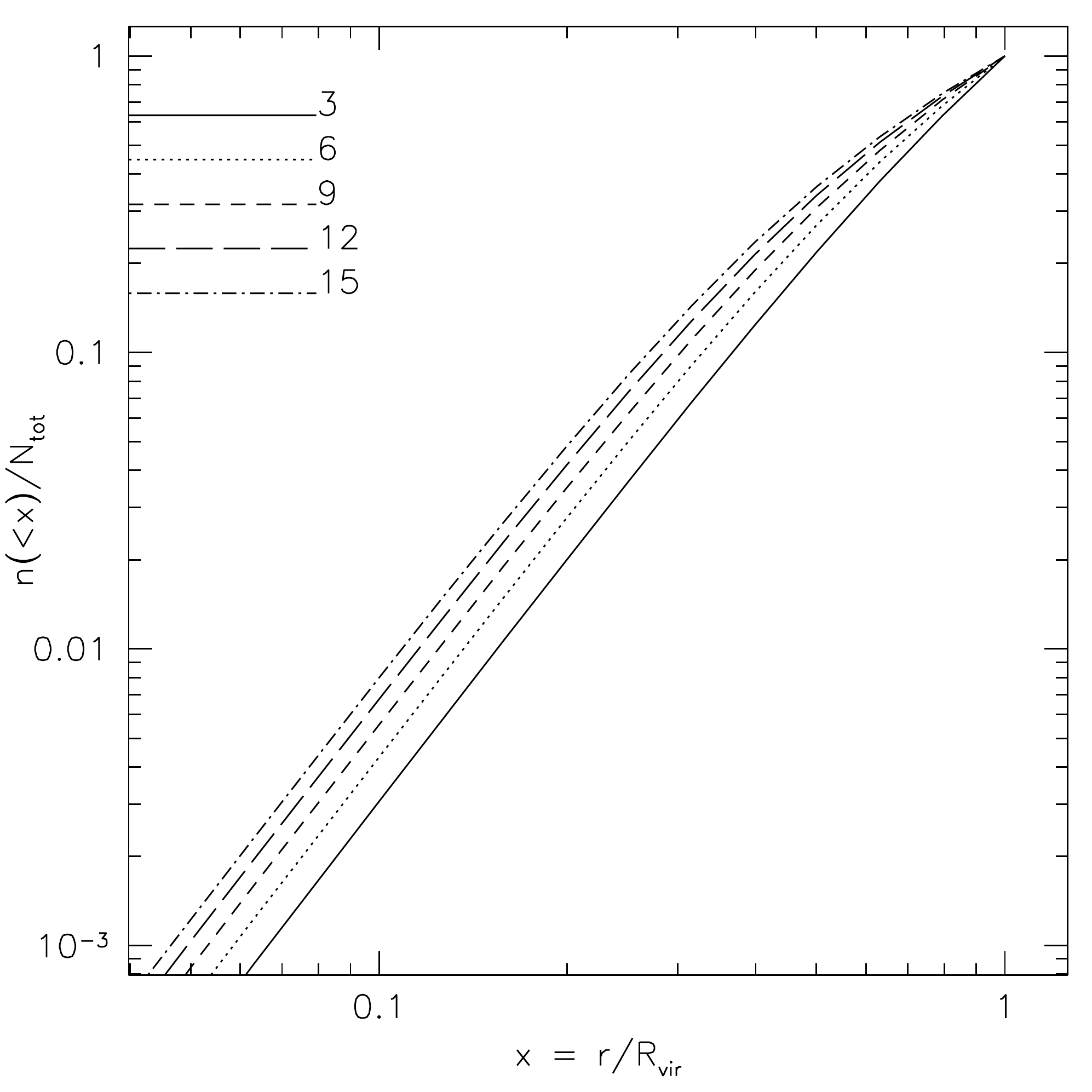}
\caption{Radial  subhalo density  distribution  model.  The  different
  curves show equation (\ref{gaoEQ})  for different values of the host
  halo concentration $c_{vir}$, as in the label.\label{gaoFIG}}
\end{figure}

\end{itemize}

\subsubsection{Dissipative baryonic component and its effects on dark matter}
Strong  lensing is  sensitive to  the matter  distribution  inside the
centre  of galaxy clusters  ($r \sim  100$ kpc).  On such  scales, the
density of  the baryonic component  becomes comparable to that  of the
dark  matter.  \citet{meneghetti03}  have  shown  that  the  brightest
central galaxy (BCG) on  strong lensing signal is generally moderately
important  for  strong lensing,  but  potentially  decisive for  those
haloes which are otherwise marginally supercritical strong lenses.

To populate haloes with a central galaxy of a certain stellar mass, we
use the Halo Occupation Distribution (HOD) technique. HOD assumes that
the stellar mass  of a galaxy is tightly correlated  with the depth of
the potential well of the halo within which it formed, thus
\begin{equation}
  M_{star} = \dfrac{2 M_{star,0}}{(M_{infall}/M_0)^{-\alpha"}+(M_{infall}/M_0)^{-\beta"}}\,,
\end{equation}
as  estimated  by  \citet{wang06},   who modelled this relation  after the
semi-analytic   galaxy  catalogue   of the   Millennium  Simulation
\citep{croton06}. In  this relation, we include a  Gaussian scatter in
$M_{star}$     at     a      given     host     halo     mass     with
$\sigma_{M_{star}}=0.148$. For the central
galaxy, we set the parameters $\alpha"=0.39$, $\beta"=1.96$, and $\log(M_{star,0)}=10.35$.

\begin{figure}
\includegraphics[width=7.5cm]{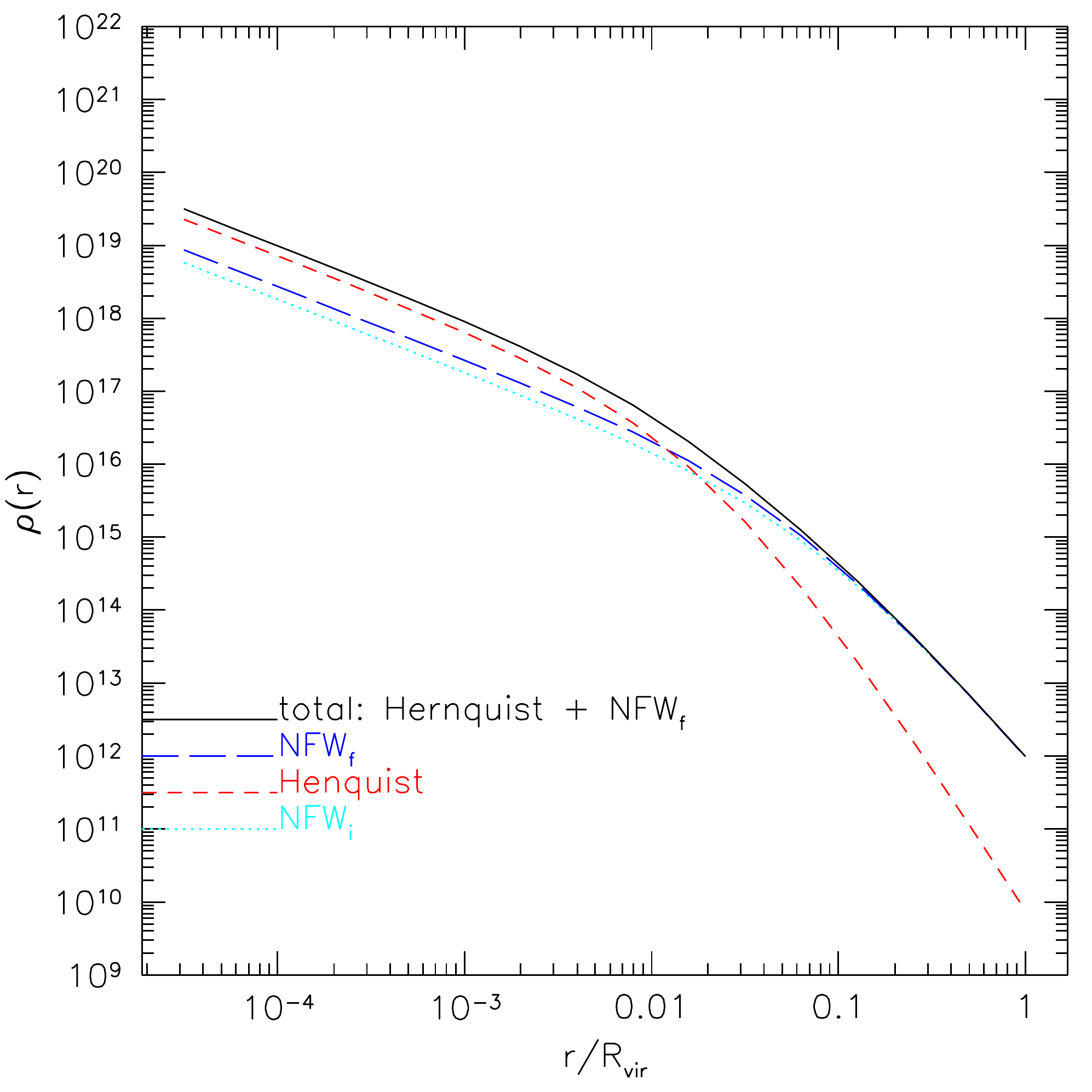}
\caption{Density    profile   of   a    halo   at    $z_l=0.25$   with
  $M_{vir}=10^{15}M_{\odot}/h$.  The   dotted  curve  shows   the  NFW
  profile,  the dashed  curve  the Henquist  profile  for the  stellar
  component of the BCG at  the halo centre with stellar mass $M_{star}
  = 5  \times 10^{12}  M_{\odot}/h$. The dissipative  baryon component
  modifies the  dark-matter profile by adiabatic  contraction as shown
  by   the  long-dashed  line.   The  solid   line  shows   the  final
  profile.\label{figdensityprofile}}
\end{figure} 

For the stellar  component of the BCG residing in  the halo centre, we
adopt the \citet{hernquist90} profile
\begin{equation}
  \rho_{star}(r) = \frac{\rho_{g}}{(r/r_g)(1+r/r_g)^3}\,. 
\end{equation}
It has  a scale radius $r_g$  related to the  half-mass (or effective)
radius  $R_e$ by $r_g  = 0.551  R_e$, as  done by  \citet{keeton01} we
define the  effective radius  to be $R_e  = 0.03 R_{vir}$.   The scale
density $\rho_g$ can be estimated  by the definition of the total mass
of a Hernquist model,
\begin{equation}
  \rho_s = \frac{M_{star}}{2 \pi r_g^3}\,.
\end{equation}

The presence  of a dissipative baryonic component  influences the dark
matter  distribution near the  host halo  centre. \citet{blumenthal86}
described  the   adiabatic  contraction  analytically,   finding  good
agreement with  numerical simulations.  The initial and  final density
profiles  -- characterized  by an  initial  radius $r_i$  and a  final
radius $r_f$, when a central galaxy is present -- are related by
\begin{equation}
  r \left[ M_{star}(r) + M_{DM,f} \right] = r_{i} M_{DM,i}(r_{i})\,,
\label{ad1}
\end{equation}
where
\begin{equation}
  M_{DM,f} = M_{DM,i} \left( 1-f_{cool}\right)\,,
\end{equation}
and $f_{cool}$ is  the baryon fraction in the halo  that cools to form
the central  galaxy. To  solve the adiabatic-contraction  equation, we
need to derive $r$ from equation (\ref{ad1}). With a Hernquist model,
\begin{equation}
  f_{cool} r^{3} + (r + r_{g})^{2}
  \left[(1-f_{cool}) r - r_{i} \right] m_{i}(r_{i}) = 0\,;
\label{eqadcontraction}
\end{equation}
This equation has a single  relevant real root. We recall that $m_{i}$
in the  previous equation defines the initial  mass profile normalized
by the  virial mass. Figure \ref{figdensityprofile}  shows the density
profile of  a halo  with virial mass  $M_{vir}=10^{15}M_{\odot}/h$ and
$c_{vir}=4$    populated   with   a    galaxy   with    stellar   mass
$M_{star}=5\times  10^{12}M_{\odot}/h$.  The  dotted  and short-dashed
curves refer  to the  initial dark matter  and stellar density  in the
halo, respectively.  The presence  of a dissipative baryonic component
contracts      the     dark     matter      distribution.      Solving
Eq.~(\ref{eqadcontraction}), the  dark matter density  profile changes
to  the long-dashed  curve. The  solid curve  shows the  total density
distribution  obtained  by  summing  the  short  and  the  long-dashed
curves. We recall that for DM-only realizations, $M_{vir}$ is given by
the sum in smooth  plus clumpy components, while $M_{vir}=M_{smooth} +
\sum_{i=1}^{N_{tot}}m_i + M_{star}$ if a BCG is present.

\subsection{Halo Lensing Properties}
\label{seclens}
In this section  we will describe how we  calculate lensing properties
starting from  the 3D matter density of  all components characterising
the  haloes. For each  component, we  project the  density on  a plane
perpendicular to the line of sight,
\begin{equation}
  \Sigma(x,y) = \int_{-\infty}^{\infty} \rho(x,y,z) \mathrm{d} z\,,
\end{equation}
where $r = \sqrt{x^2+y^2+z^2}$ with the coordinate origin put into the
host halo centre, and defining $\xi=\sqrt{x^2+y^2}$.

\begin{figure}
\includegraphics[width=7.5cm]{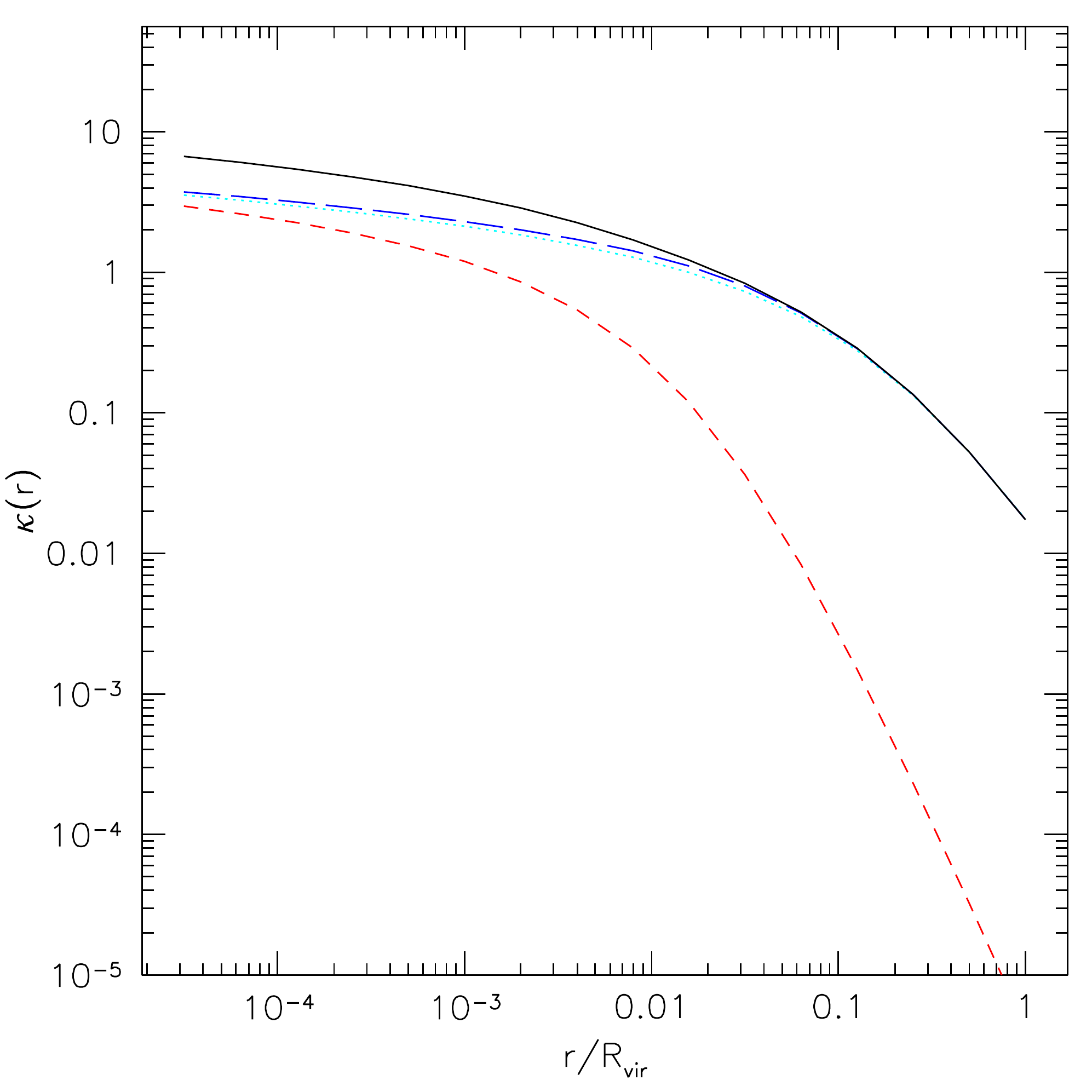}
\caption{Convergence profile of the  same cluster, assuming sources at
  redshift $z_s=2$. \label{figkappaprofile}}
\end{figure}

The projected mass density of  the NFW, Hernquist and SIS profiles can
be given analytically \citep{bartelmann96,bartelmann01,keeton01}:
\begin{eqnarray}
  \Sigma_{NFW}(x,y)= \frac{2 \rho_s r_s}{\zeta^2 -1 } F(\zeta)\,, \\
  \Sigma_{star}(x,y) = \rho_g r_g \frac{(2+\eta^2) G(\eta) - 3}{(\eta^2 - 1)^2}\,, \\
  \Sigma_{sub}(x,y) = \frac{\sigma_v}{2 G \xi}\,,
\end{eqnarray}
where $\zeta\equiv\xi/r_s$, $\eta\equiv\xi/r_g$, and the two functions
$F(\zeta)$ and $G(\eta)$ are
\begin{displaymath}
  F(\zeta) = \left\{ \begin{array}{ll}
    1- \frac{2}{\sqrt{\zeta^2-1}} \arctan\sqrt{\dfrac{\zeta-1}{\zeta+1}}&\zeta>1, \\
    1- \frac{2}{\sqrt{1-\zeta^2}}
    \mathrm{arctanh}\sqrt{-\dfrac{\zeta-1}{\zeta+1}} &\zeta<1, \\
    0& \zeta=1;\\
  \end{array} \right.
\end{displaymath}

\begin{displaymath}
  G(\eta) = \left\{ \begin{array}{ll}
    \frac{1}{\sqrt{\eta^2-1}} \tan^{-1}\sqrt{\eta^2-1} &\eta>1, \\
    \frac{1}{\sqrt{1-\eta^2}} \tanh^{-1}\sqrt{1-\eta^2} &\eta<1, \\
    1& \eta=1.\\
  \end{array} \right.
\end{displaymath}
We recall that the subhalo  density profile is truncated at the radius
enclosing its total (bound) mass. The convergence is the appropriately
scaled surface mass density $\Sigma(x,y)$,
\begin{equation}
  \kappa(x,y) = \frac{\Sigma(x,y)}{\Sigma_{cr}}\,
\end{equation}
where
\begin{equation}
  \Sigma_{cr} = \frac{c^2}{4 \pi G} \frac{D_s}{D_l D_{ls}}\,,
\end{equation}
is the critical surface  mass density, containing the angular-diametre
distance $D_l$, $D_s$  and $D_{ls}$ from the observer  to the lens, to
the source, and from the lens to the source, respectively.

\begin{figure*}
\includegraphics[width=\hsize]{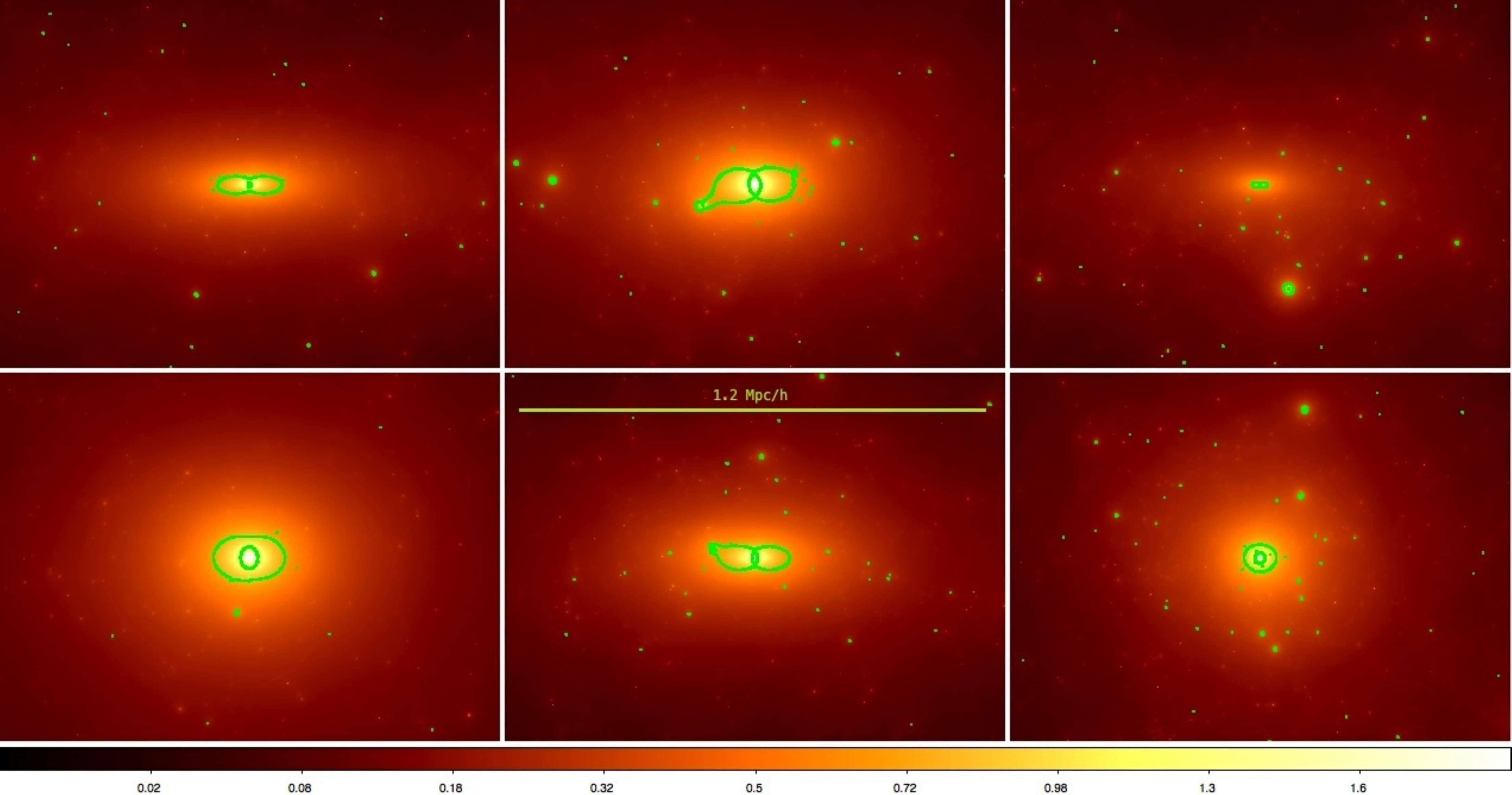}
\caption{Convergence  maps of  cluster-sized haloes  created  with our
  \texttt{MOKA} code at redshift $z_{l}=0.25$, assuming sources at the
  single redshift  $z_s=2$. The  solid curves indicate  the tangential
  and  radial critical  curves.   All  haloes have  a  virial mass  of
  $10^{15}M_{\odot}/h$    and    a    minimum    subhalo    mass    of
  $10^{10}\,M_{\odot}/h$. \label{figclusters}}
\end{figure*}

Figure  \ref{figkappaprofile} shows the  convergence profiles  for the
halo  components  presented  in Figure  \ref{figdensityprofile},  with
sources assumed at $z_s=2$.

For each dark-matter halo, the total convergence map is the sum of all
contribution maps,
\begin{eqnarray}
\label{eqkappatot}
  \kappa(x,y) = \kappa_{DM}(x,y) &+&\\ 
  \kappa_{star}(x,y) &+& \sum_{i=1}^{N}\kappa_{sub,i}(x-x_c,y-y_c)\,,\nonumber 
\end{eqnarray}
where  $x_c$ and  $y_c$ represent  the center  of mass  of  the $i$-th
substructure. To introduce ellipticity  into our model, we draw random
numbers  from the  \citet{jing02} distributions  for the  axial ratios
$a/b$ and  $a/c$, requiring $a  b c =  1$.  Once the axial  ratios are
known, we deform the  convergence map accordingly.  To randomly orient
the  halo, we  choose a  random point  on a  sphere identified  by its
azimuthal and elevation angles and  rotate the halo ellipsoid by these
angles.   We  assign the  same  projected  ellipticity  to the  smooth
component, to the  stellar density and to the  subhalo spatial density
distributions. For  the dark matter density  distribution in subhaloes
we  assume  spherically symmetric  models,  this  because the  subhalo
typical  scale is  much smaller  than the  virial radius  of  the host
system in  which they are  located.  Having elliptical  subhaloes will
impact slightly  on the total  strong lensing cross-section.   We have
chosen  to assign  the same  ellipticity to  the BCG  and to  the dark
matter  density distribution  in  the host  following  the results  of
\citet{fasano10} who found that the  shape of the BCG tends to reflect
that  of the  associated dark  matter halo.   Figure \ref{figclusters}
show the  convergence maps of  six galaxy clusters generated  with our
algorithm.  All haloes are located  at redshift $z=0.25$ and possess a
virial  mass equal  to  $10^{15}M_{\odot}/h$. The  source redshift  is
$z_s=2$.  The subhalo mass resolution is $10^{10}\,M_{\odot}/h$.  This
ensures    a    substructure    mass    fraction    compatible    with
\citet{richard10}, and  consistent with the  fact that systems  with a
lower substructure fraction  tend to form at higher  redshift, and are
thus more concentrated \citep{smith08}.

From  the convergence  map,  the effective  potential  and the  scaled
deflection angle can be obtained by means of the equations
\begin{equation}
  \Phi(x,y) = \frac{1}{\pi} \int_{\mathbb{R}^2} \kappa(\vec{\xi}')
  \ln |\vec{\xi}-\vec{\xi}'| \mathrm{d}^2 \xi',
\end{equation}
and
\begin{equation}
  \alpha(x,y) = \frac{1}{\pi} \int_{\mathbb{R}^2} \kappa(\vec{\xi}')
  \frac{\vec{\xi}-\vec{\xi}'}{|\vec{\xi}-\vec{\xi}'|} \mathrm{d}^2 \xi'.
\end{equation}
A source will  be seen at the angular  position $\vec{\theta}$ related
to its intrinsic angular position $\vec{\beta}$ by the lens equation
\begin{equation}
  \vec{\beta} = \vec{\theta} - \vec{\alpha}(\vec{\theta}).
\end{equation}
Derivatives of the lensing potential are denoted by subscripts,
\begin{equation}
  \frac{\partial^2 \Phi(x,y)}{\partial \xi_i \partial \xi_j} \equiv \Phi_{ij},
\end{equation}
where $\xi_i=x$ when $i=1$ and  $\xi_i=y$ when $i=2$. We introduce the
pseudo-vector field of the shear $\vec{\gamma}=(\gamma_1,\gamma_2)$ by
its components
\begin{eqnarray}
  \gamma_1(x,y) &=& \frac{1}{2} ( \Phi_{11} - \Phi_{22} ), \\
  \gamma_2(x,y) &=& \Phi_{12} = \Phi_{21}. 
\end{eqnarray}

Light  bundles  are deflected  differentially  and  thus distorted  as
described by the Jacobian matrix
\begin{equation}
  A = \left( \delta_{ij} - \Phi_{ij}\right),
\end{equation}
with the eigenvalues
\begin{eqnarray}
  \lambda_t &=& 1 - \kappa - \gamma, \\
  \lambda_r &=& 1 - \kappa + \gamma.
\end{eqnarray}
The conditions $\lambda_t=0$ and  $\lambda_r=0$ define the location of
the tangential and radial critical curves in the lens plane, where the
magnification is  formally infinite.  Mapping the critical  curve back
into the  source plane by the  lens equation gives  the tangential and
radial caustics.

\begin{figure*}
\includegraphics[width=7.5cm]{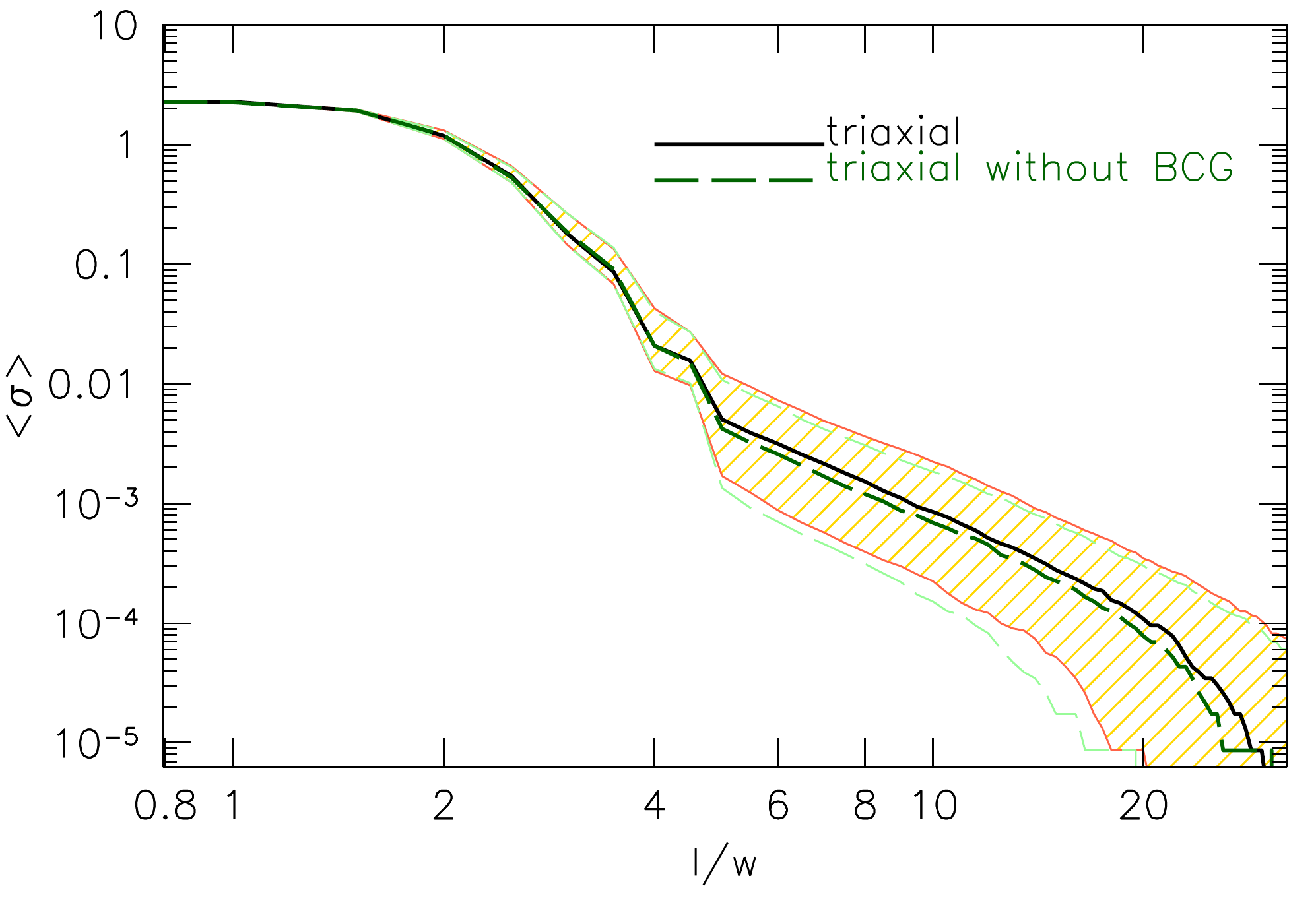}
\includegraphics[width=7.5cm]{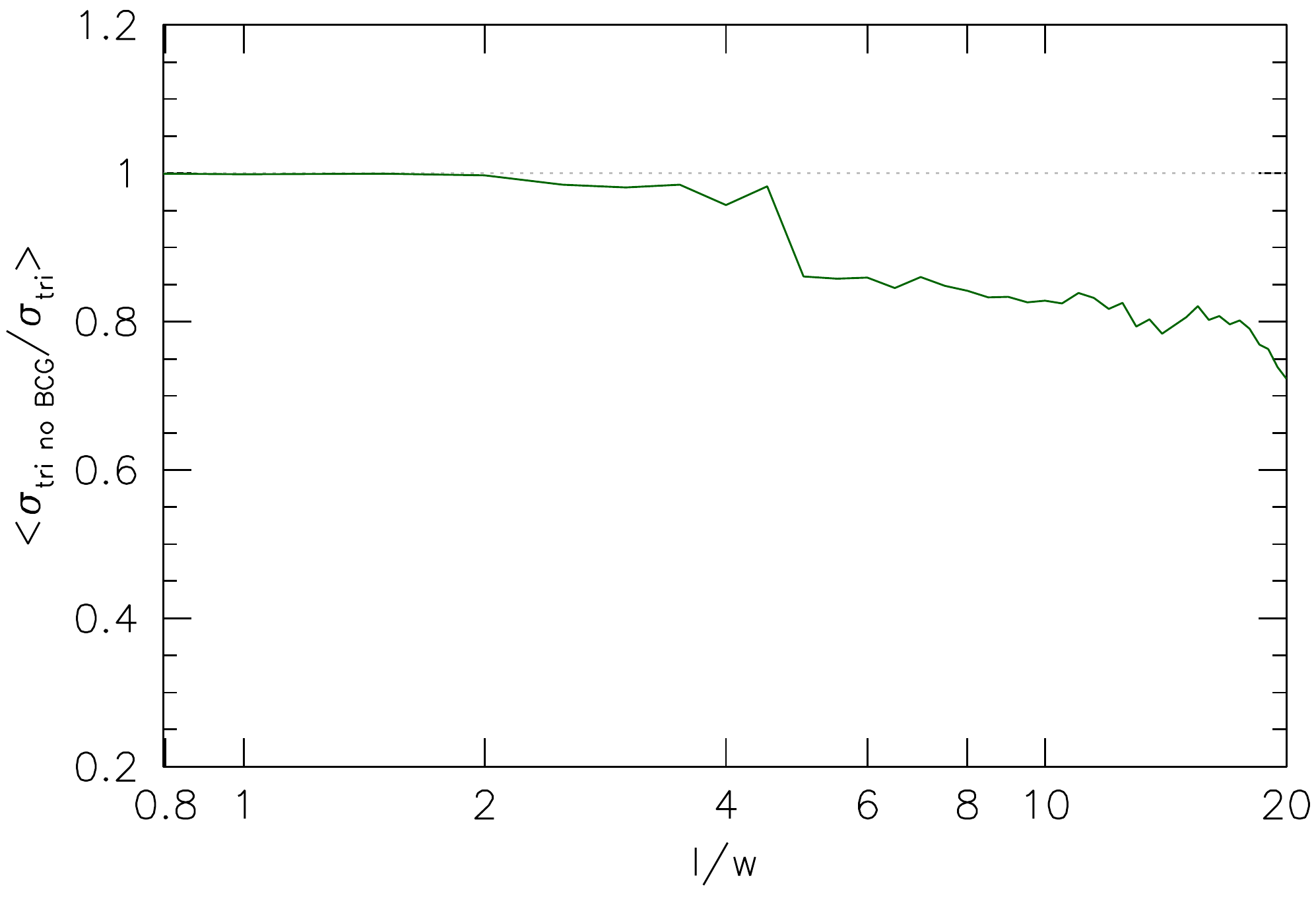}
\caption{Left  panel:  median of  the  strong  lensing cross  sections
  $\sigma$ of  a sample of simulated  haloes with $10^{15}M_{\odot}/h$
  at $z_l=0.25$  with sources at $z_s=2$  as a function  of $l/w$. The
  solid curves  show the median and $25-75\%$  percentiles of $\sigma$
  for triaxial haloes with a BCG, while the dashed curves are the same
  sample but with only DM components. Right panel: median ratio of the
  strong lensing  cross section  without and with  BCG as  function of
  $l/w$.    The   scatter   for   $l/w>5$   is   of   the   order   of
  $0.4$. \label{figwBCG}}
\end{figure*}
Table~\ref{tabPar} summarizes the list of parameters controlling
\texttt{MOKA}.

\begin{table*}
\caption{ \label{tabPar} List of parameters present in \texttt{MOKA
    1.0}:
\href{http://cgiocoli.wordpress.com/research-interests/moka}{http://cgiocoli.wordpress.com/research-interests/moka}.}
\begin{tabular}{ l |  r | }
 \hline                       
  Halo concentration: & drawn from a log-normal distribution around the
  mean value at fixed mass$^1$ \\ 
  Halo virial radius: & defined from the spherical collapse model \\ 
  Axial ratios: &  drawn from the \citet{jing02} model$^2$ \\
  Halo orientation: &  random picking the point on a sphere  \\
  Subhalo mass function: & random sampling the \citet{giocoli10} model \\
  Subhalo distribution: & random sampling the \citet{gao04} model,
  also NFW model has been implemented \\
  Subhalo velocity dispersion: & fixed by the equation
  (\ref{EQsigma}) \\
  BCG stellar mass: & sampling the \citet{wang06} model with a
  gaussian scatter \\
  BCG effective radius: & fixed to be the $4\%$ of the host halo
  virial radius \citep{keeton01} \\
  \hline  
\end{tabular}
\begin{flushleft}
$^1$\texttt{MOKA} is flexible being able to work with different
  mass-concentration relation model \\
$^2$ellipticity can be turned off
\end{flushleft}
\end{table*}

\section{Lens structural properties and strong lensing signals}
\label{secresults1}
In this section  we discuss the strong lensing  efficiency of clusters
produced by \texttt{MOKA}.  Most of  the discussion is focussed on the
lensing cross section  $\sigma$. This is defined as  the region on the
source  plane from where  the sources  are mapped  into images  with a
certain length-to-width ration $l/w$.

For each cluster  we create a high-resolution deflection  angle map of
$2048\times  2048$ pixels  centred on  the cluster  centre,  with side
length equal to  the virial radius. From this  we numerically estimate
the   lensing  cross   section   by  means   of  ray-tracing   methods
\citep{meneghetti05a,meneghetti05b,fedeli06,meneghetti11}.  A  typical
calculation  takes  about  one  minute  on a  $3.06$  GHz  single-core
processor.   Starting from  the observer,  bundles of  light  rays are
traced  back to  the source  plane. Using  an adaptive  grid,  this is
populated  by  elliptical sources  of  a  fixed  equivalent radius  of
$0.5''$.   The  number of  highly  magnified  images  is increased  by
refining  the spatial resolution  on the  source plane  near caustics.
Analysing  the images  individually and  measuring  their length-width
ratio $l/w$,  we can define the  lensing cross section  for giant arcs
$\sigma_{l/w}$  as the  region  on  the source  plane  from where  the
sources are mapped  into images with a certain  $l/w$.  Giant arcs are
commonly  defined as  distorted  images with  a length-to-width  ratio
$l/w\geq7.5$  or $l/w\geq10$.  We  shall discuss  our results  for the
three different values $l/w=5$, $7.5$ and $10$.

The integral of  the strong lensing cross section  of a galaxy cluster
for  different  $z_s$, weighted  by  a  source redshift  distribution,
allows to  quantify the  number of gravitational  arcs expected  to be
seen. The strong lensing analysis of galaxy cluster size haloes from a
$\mathrm{\Lambda}$CDM  numerical  simulations by  \citet{bartelmann98}
has revealed  that those  haloes produce an  order of  magnitude fewer
gravitational arcs  than observed.  Studying how  strong lensing cross
sections change with structural halo properties may help to understand
possible  current limitations  of numerical  simulations,  without the
need   of   advocating    cosmological   models   with   dark   energy
\citep{bartelmann03}.

Recent observational studies of  strong lensing clusters revealed that
observed  clusters  seem  to   be  more  concentrated  than  simulated
clusters. This  causes more prominent strong lensing  features. In the
following, we shall  quantify how much the presence of  a BCG and halo
triaxiality  influence  the strong  lensing  cross  section for  giant
arcs. We  have created a  sample of $128$  high-resolution convergence
maps of haloes with mass $10^{15}\,M_{\odot}/h$ and sources located at
a  fixed redshift  $z_s=2$. This  study is  considered as  a reference
starting point to present our code \texttt{MOKA}.  At the present time
we   are  performing   a   statistical  analysis   of   haloes  in   a
$\mathrm{\Lambda}CDM$  universe  to give  an  estimate  of the  strong
lensing  cross section  as  a function  of  redshift and  of the  halo
abundance (Boldrin et al. in preparation).
 
\subsection{BCG}
Using  a  variety  of  Gas-dynamical  simulations,  \citet{puchwein05}
studied the  impact of  the gas component  on strong  lensing signals,
finding that  the formation of stars  mainly in the  central region of
the cluster  tends to increase  the cross section by  about $20-30\%$,
depending on  lens redshift.  To consider an  analogous case,  we have
generated  a sample  of  haloes with  the  same structural  properties
($M_{vir}=10^{15}M_{\odot}/h$,  $z_l=0.25$   and  $z_s=2$),  with  and
without a central  galaxy. We recall that for  pure dark matter lenses
the virial  mass takes into account  the sum of the  smooth plus clump
components, while in simulations with BCG the total $M_{vir}$ includes
also  the presence of  the stellar  system. In  both cases,  the total
masses  are identical.  For  each halo  we  have post-processed  their
deflection angle  map and estimated the strong  lensing cross sections
as a  function of arc length-to-width  ratio $l/w$. The  left panel of
Figure \ref{figwBCG} shows the  median strong lensing cross section as
a function of $l/w$ for a  sample of clusters with (solid) and without
BCG  (dashed  curve). The  shaded  regions  enclose  $25-75\%$ of  the
data. The  right panel of the  same figure shows for  each cluster the
ratio   of  the   strong  lensing   cross  section   as   function  of
length-to-width ratio with and  without a BCG, including the estimated
median. The Figure confirms that  at redshift $z_l=0.25$, the value of
$\sigma$ for arcs with length-to-width ratio $l/w>5$ tends to be lower
of a $20\%$ when the BCG is not included.

\begin{figure*}
\includegraphics[width=7.5cm]{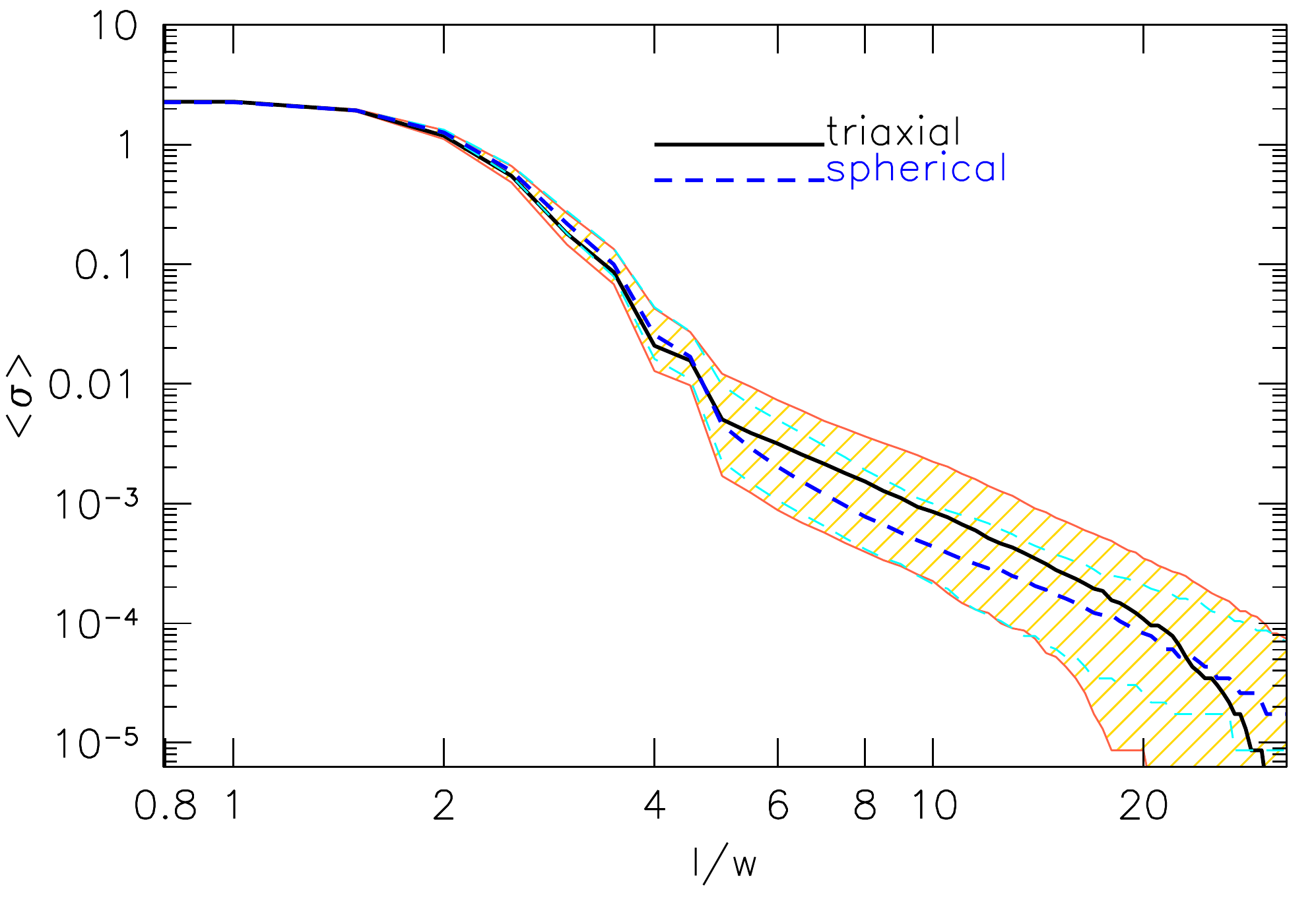}
\includegraphics[width=7.5cm]{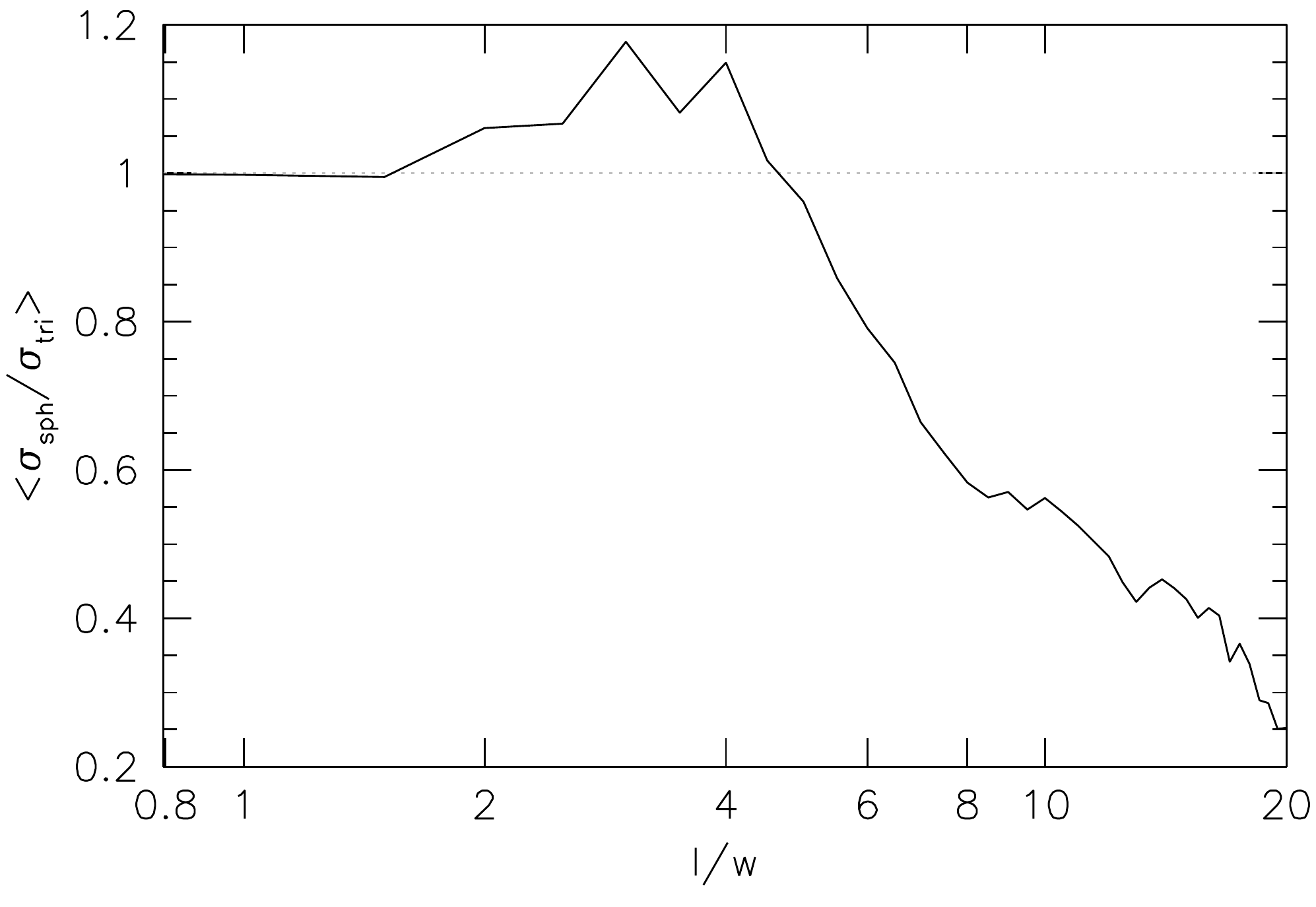}
\caption{Left  panel:  median of  the  strong  lensing cross  sections
  $\sigma$ of  a sample of simulated  haloes with $10^{15}M_{\odot}/h$
  at  $z_l=0.25$ with  sources at  $z_s=2$ as  function of  $l/w$. The
  solid  line  shows  the  median  and the  $25-75\%$  percentiles  of
  $\sigma$ for triaxial haloes (same as Fig.~\ref{figwBCG}), while the
  dashed  line is  for  spherical  haloes. BCGs  are  present in  both
  samples.  Right  panel: median  of the ratio  of the  strong lensing
  cross section  for the trixial  and spherical haloes as  function of
  $l/w$. Neglecting  triaxiality decreases  the singal for  $l/w>5$ by
  $20-70\%$.    The  scatter   for  $l/w>5$   is  of   the   order  of
  $0.4$. \label{figwsph}}
\end{figure*}

\subsection{Triaxiality}
Now we consider the case of spherical and triaxial haloes. Triaxiality
affects strong lensing through  the halo orientation.  A cluster whose
major  axis  is oriented  along  the  line of  sight  will  be a  more
efficient  strong  lens  than  if  it  is  oriented  otherwise.   This
orientation  bias  influences also  mass  and concentration  estimates
\citep{meneghetti10b}.  To  quantify how important  triaxiality is for
the appearance  of giant arcs, we  have created a  sample of spherical
haloes and  compared their strong  lensing cross sections to  those of
our  fiducial sample  of  triaxial haloes  whose  mean ellipticity  is
$\epsilon_{3D}=0.15$\footnote{We   define  the   3D   ellipticity  as:
  $\epsilon_{3D}\equiv(c-a)/[2(a+b+c)]$, where $a$, $b$ and $c$ define
  the   minor,   median  and   major   axes  respectively.}.    Figure
\ref{figwsph} shows the strong lensing  cross section as a function of
$l/w$   for  the  same   sample  of   either  triaxial   or  spherical
galaxy-cluster  haloes. Both  samples contain  BCGs.  The  right panel
shows  the median  ratio  of  $\sigma$ between  the  two samples:  for
$l/w>5$,  the  strong-lensing  signal  tends to  decrease  by  between
$20\,\%$ and  $70\,\%$ in spherical compared to  triaxial haloes. This
confirms results obtained  with a small sample of  simulated haloes by
\citet{meneghetti07b}.    Notice    also   that   arcs    with   small
length-to-width ratios (between  $2$ to and $5$) are  more abundant in
spherical haloes because of their relatively lower shear.

\subsection{Scatter}
\begin{figure}
\includegraphics[width=\hsize]{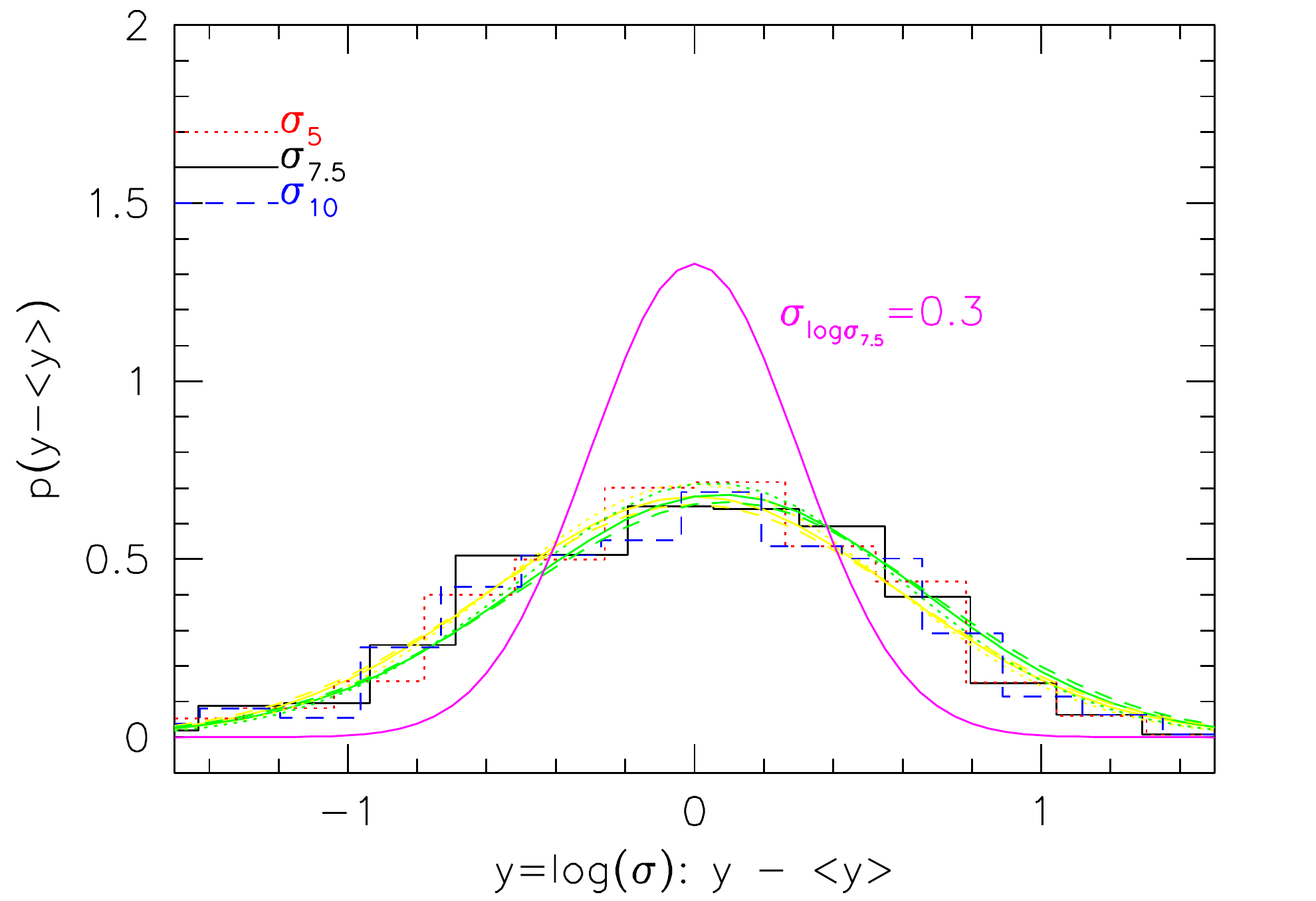}
\caption{Distribution of the strong  lensing cross section at $l/w=5$,
  $7.5$ and $10$. To fit  the distributions, we have considered both a
  Gaussian and an Edgeworth expansion, for $y=\log(\sigma)$, which are
  superposed for corresponding value of $\sigma_{l/w}$. The solid line
  shows a Gaussian with $\sigma=0.3$.  \label{figdistsigma}}
\end{figure}

At a  given host  halo mass and  lensing redshift, the  strong lensing
cross section scatters  around its mean.  This scatter  is due to both
halo  triaxiality,  presence  and  distribution of  substructures  and
concentrations.      \citet{fedeli10}    fit     the     scatter    in
$y=\log(\sigma_{7.5})$ for galaxy cluster sized haloes with a Gaussian
distribution, or an Edgeworth expansion around it, including skewness.
Figure \ref{figdistsigma} shows the distribution of the strong lensing
cross  sections for  three length-to-width  ratios $l/w=5$,  $7.5$ and
$10$. In all  cases, Gaussian distributions fit the  data well with no
significant   deviation  from  the   Edgeworth  expansion,   which  we
superpose.    The    distributions   have   standard    deviation   of
$\sigma_{\log(\sigma)}$  slightly  increasing   with  $l/w$;  we  find
$\sigma_{\log(\sigma)}=0.56$, $0.59$ and $0.61$ for $l/w=5$, $7.5$ and
$10$, respectively.  These values  are approximately twice those found
by \citet{fedeli10}.  We  believe that the main reason  for the larger
standard  deviation is  how the  halo boundaries  were defined  in the
simulations and with our algorithm: while we include all matter of the
main  halo along the  line of  sight, \citet{fedeli10}  have extracted
boxes  of   $5$  Mpc  side  length  from   their  simulation.  Another
contribution to  this difference  is given by  the fact that  for each
halo  \citet{fedeli10} estimate  the strong  lensing cross  section as
mean  of the three  projections: at  the end  this procedure  cuts the
wings of their distribution.

\subsection{Concentration}
Haloes of the same mass have different concentrations reflecting their
different assembly  histories. We  show in Appendix  \ref{appcrel} how
the  \textit{rms} scatter  in  the strong  lensing  cross sections  is
related to  the concentration scatter  and thus to the  different host
halo merger histories.

\begin{figure}
\includegraphics[width=\hsize]{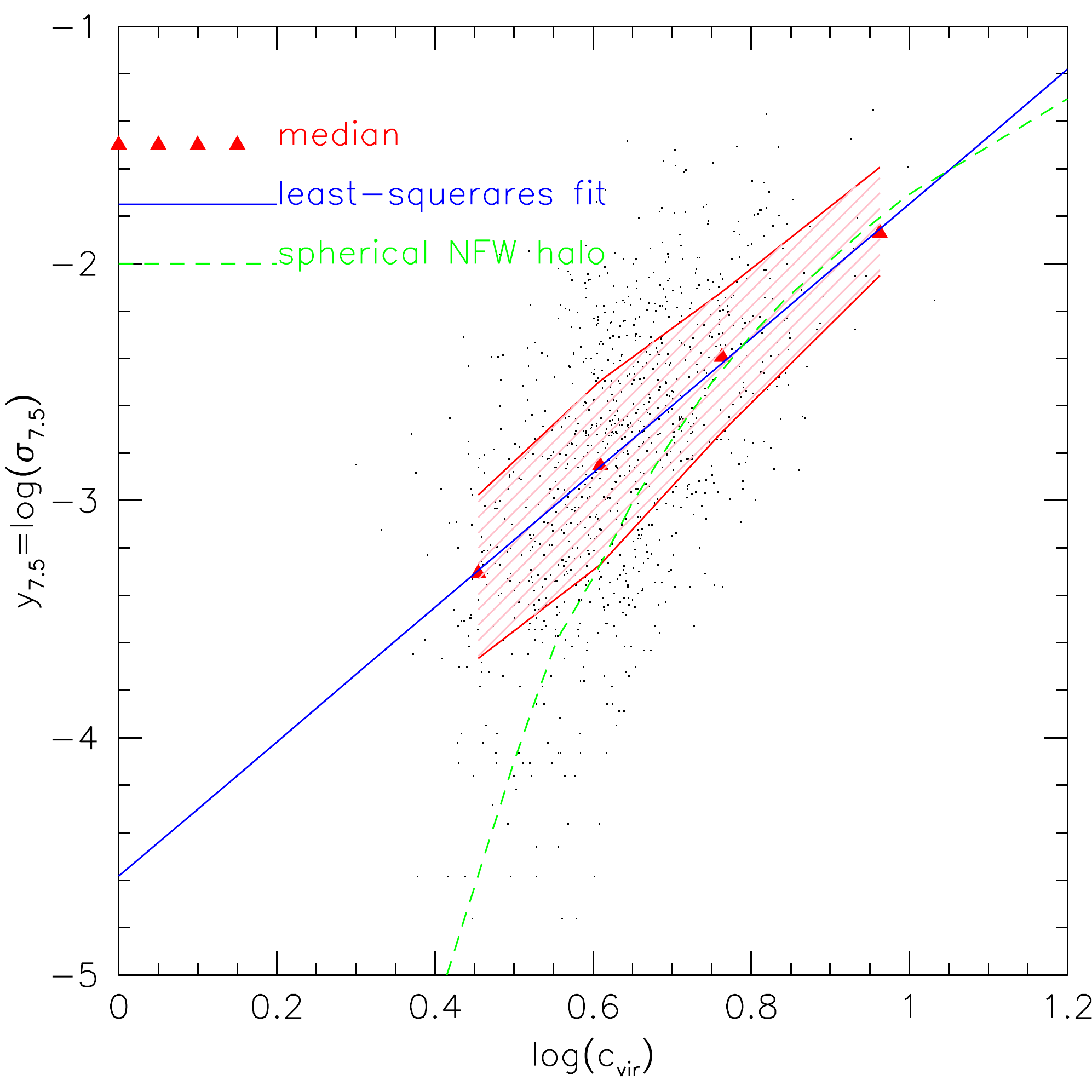}
\caption{Correlation between the strong lensing cross section for arcs
  with   length-to-width   ratio    $\ge7.5$   and   the   host   halo
  concentration. Filled  triangles show the median  of the correlation
  and the shaded region encloses $25-75\%$ of the data. The solid line
  is a least-squares fit to  these data points whose slope is $2.83\pm
  0.06$.   The dashed  curve shows  the prediction  for  spherical and
  smooth  NFW haloes (without  a central  galaxy).\label{figsigmac} In
  this case  the sample of  haloes is made  up of $1024$  systems with
  mass $M_{vir}=10^{15}\,M_{\odot}/h$ at redshift $z_l=0.25$.}
\end{figure}

Strong lensing  depends mainly  on the matter  density in  the cluster
cores, where the dark-matter  density profile approaches a logarithmic
slope near $-1$. The scale where the logarithmic density slope is $-2$
defines  the scale  radius $r_s$  and the  host halo  concentration by
$c_{vir}=r_s/R_{vir}$.  Higher   concentrations  cause  larger  strong
lensing cross sections. Figure \ref{figsigmac} shows this relation for
our halo sample at $z_l=0.25$. Filled triangles show the median of the
relation, with  error bars enclosing the quartile  of the distribution
in each $\sigma_{7.5}$  bin. The solid line shows  a least-squares fit
to the data whose slope is  $2.83\pm 0.06$. The dashed curve shows the
predicted relation  between the strong  lensing cross section  and the
host  halo concentration  for spherical  smooth NFW  haloes  with mass
$10^{15}M_{\odot}/h$. Notice  that our  simulated lenses tend  to have
larger strong lensing cross  sections at low concentration compared to
spherical NFW haloes.  This is mainly due to  the BCGs and triaxiality
which both tend to increase the projected central mass density.

\subsubsection{Statistical Properties}
To  summarize, we  discuss  now  the strong  lensing  efficiency of  a
population of clusters generated  with \texttt{MOKA} and we compare it
to that  of a  similar sample of  numerically simulated  cluster taken
from a cosmological box. 

\begin{figure}
\includegraphics[width=\hsize]{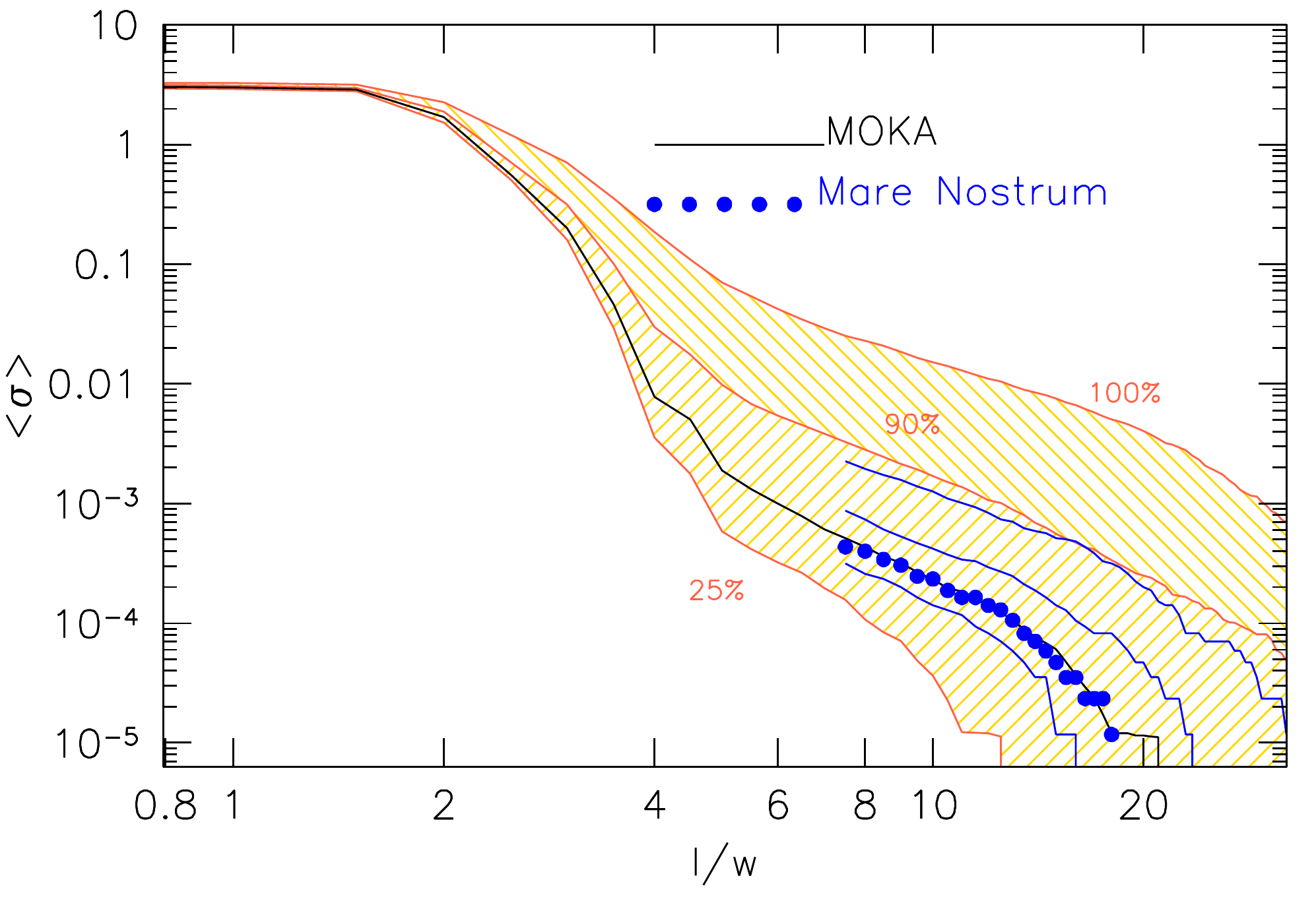}
\caption{Comparison  between the median  strong lensing  cross section
  $\sigma$ as  a function of  $l/w$ measured in  analytically modelled
  haloes  (solid line)  and in  haloes from  the  \texttt{MARE NOSTRUM
    UNIVERSE} (filled circles) for a sample of cluster-sized haloes in
  the mass range $6-7  \times 10^{14}M_{\odot}/h$.  The shaded regions
  enclose $25\%-90\%$  and $90\%-100\%$ for a  sample of \texttt{MOKA}
  haloes, while the three solid curves illustrate the same regions for
  the haloes in the \texttt{MARE NOSTRUM UNIVERSE}.  \label{figwMN}}
\end{figure}

\citet{meneghetti10a} studied the strong lensing by a sample of galaxy
clusters   from  the  \texttt{MARE   NOSTRUM  UNIVERSE}.    In  Figure
\ref{figwMN}, we compare our findings with the median cross section of
a  sample  of  simulated   haloes  with  masses  between  $6-7  \times
10^{14}M_{\odot}/h$ at redshift $z_l=0.18$ (filled circle).  Since the
\texttt{MARE NOSTRUM} simulation  does not include radiative processes
and star  formation, haloes  formed therein do  not contain  a central
massive galaxy. To make them compatible with the \texttt{MOKA} haloes,
we add  a BCG with  mean $10^{12}\,M_{\odot}/h$ at their  center.  The
solid  line shows  the median  cross section  (shaded  regions enclose
$25-90\%$ and  $90-100\%$ of the  data) for a sample  of \texttt{MOKA}
haloes in  the same mass range,  whose abundance has  been weighted by
the \citet{sheth99b}  mass function.  The three solid  curves show the
same  regions for the  haloes in  the \texttt{MARE  NOSTRUM UNIVERSE}.
Our algorithm well  reproduces the median of the  strong lensing cross
sections measured  in the  numerical simulation. It  doesn't reproduce
the scatter  of the lensing  cross sections for the  reasons discussed
above.

\subsubsection{Host halo concentration}
Recent studies  combining strong and  weak lensing of  galaxy clusters
have not  only disagreed with predictions  of the number  of arcs, but
the observed  clusters also  seem to have  Einstein radii  larger than
predicted     in      the     $\mathrm{\Lambda     CDM}$     cosmology
\citep{zitrin11a,zitrin11c}. This may emphasise that observed clusters
are  more  concentrated  than  those found  in  numerical  simulations
\citep{oguri05,oguri09}.  However,   over-concentrations  in  observed
haloes may also result from an orientation bias \citep{meneghetti11}.

We shall now quantify by how much the strong lensing signal changes if
the normalization of the  mass-concentration relation is increased. We
perform different  simulations of the same halo  sample, assuming that
the mass-concentration relation  is given by Eq.~(\ref{eqzhao}), where
we include a normalization factor $c_0$. A reasonable choice for $c_0$
is  a value  between $-3.2$  and $6.4$,  consistently  with simulation
results for the mean concentration scatter at fixed halo mass.

\begin{figure}
\includegraphics[width=\hsize]{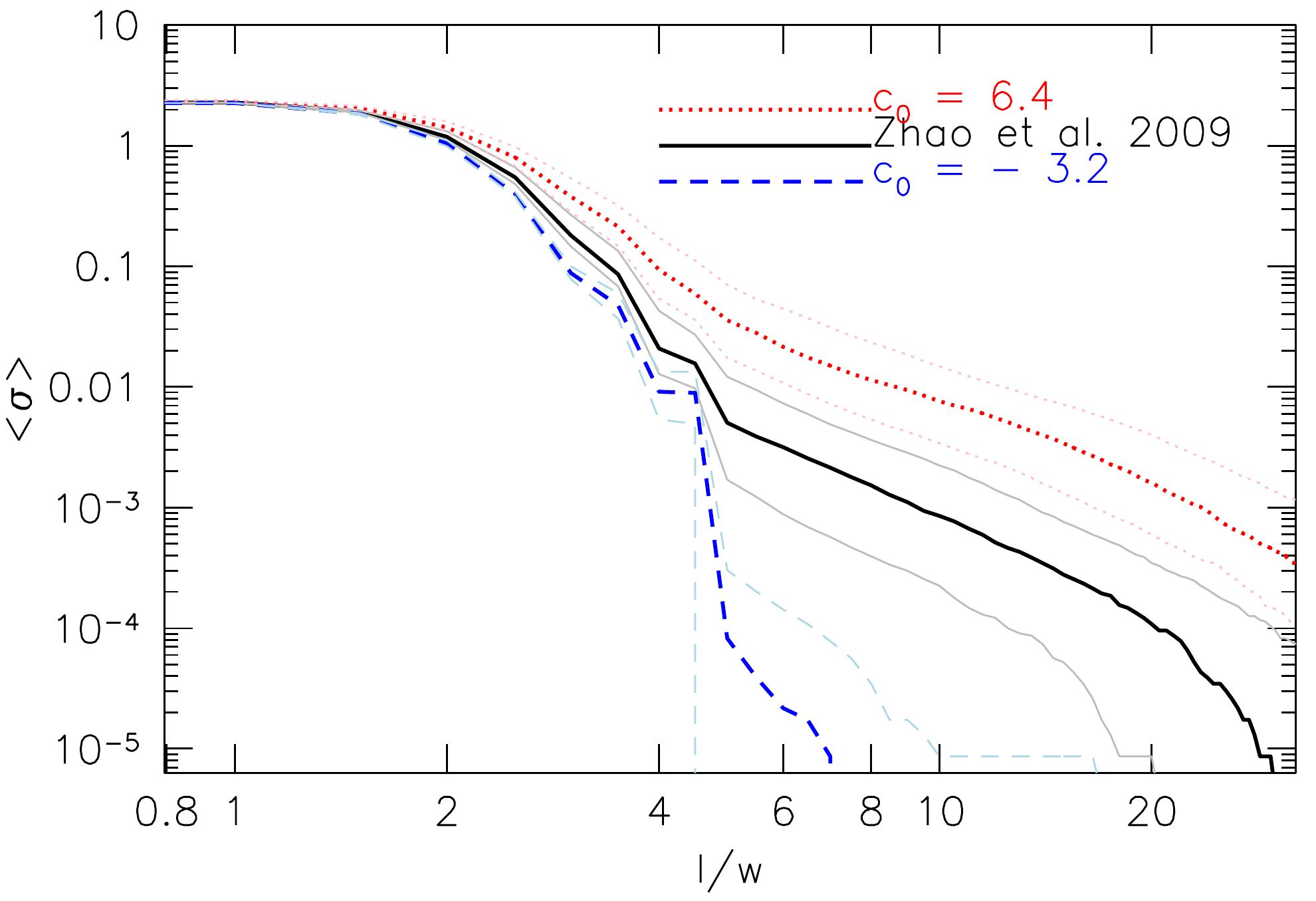}
\caption{Dependence of the median  strong lensing cross section on the
  normalization  of the  mass-concentration relation.  The  solid line
  shows the median cross section for our fiducial sample of cluster at
  $z_l=0.25$, the  dotted and the  dashed lines show the  relation for
  higher   and    lower   value   of   the    normalization   in   the
  mass-concentration                 relation                relation:
  $c_{vir}'(M_{vir},z_l)=c_{vir}(M_{vir},z_l)+c_0$, for \citet{zhao09}
  $c_0=0$.\label{figsigmanc}}
\end{figure}

\begin{figure*}
\includegraphics[width=5.5cm]{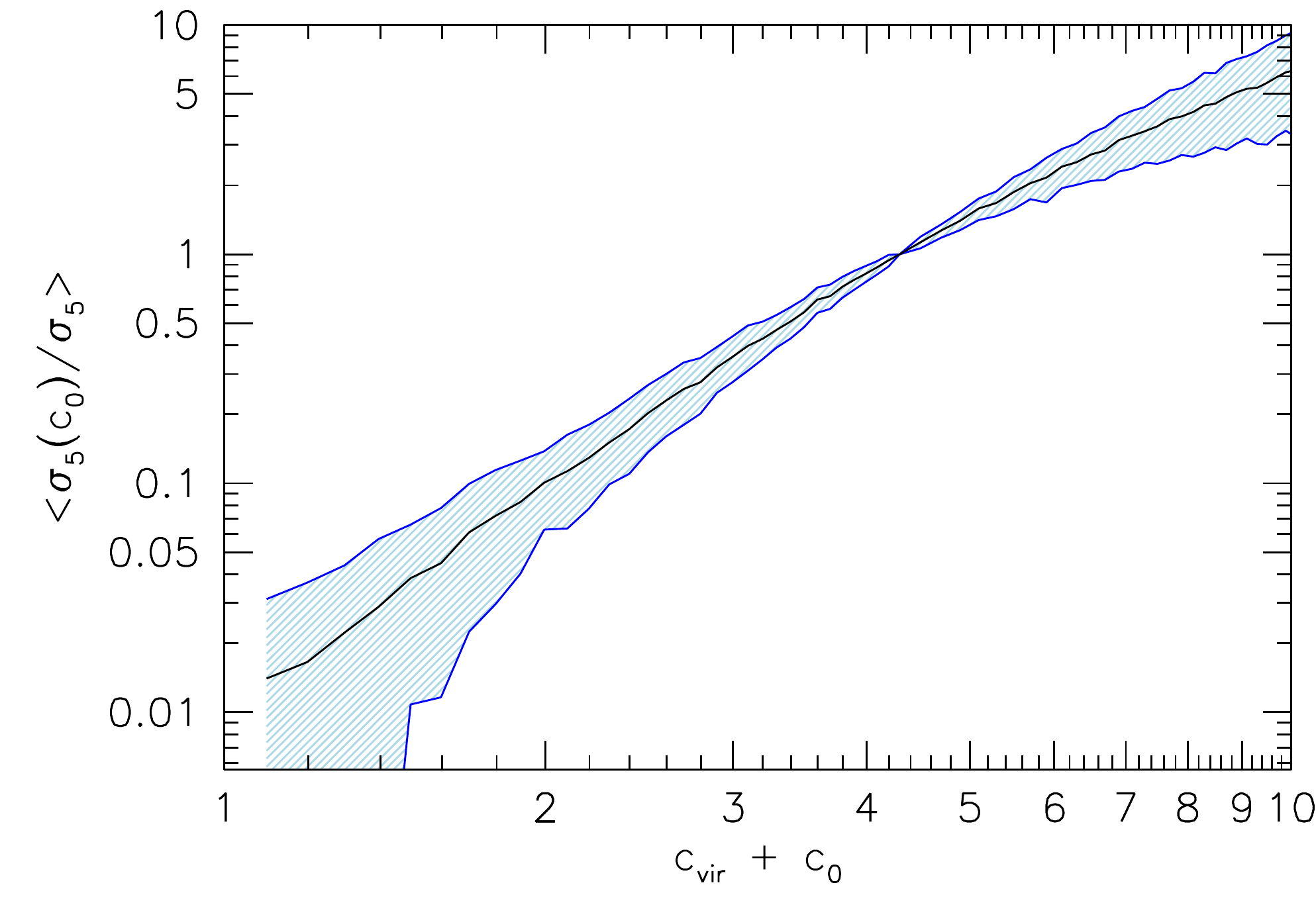}
\includegraphics[width=5.5cm]{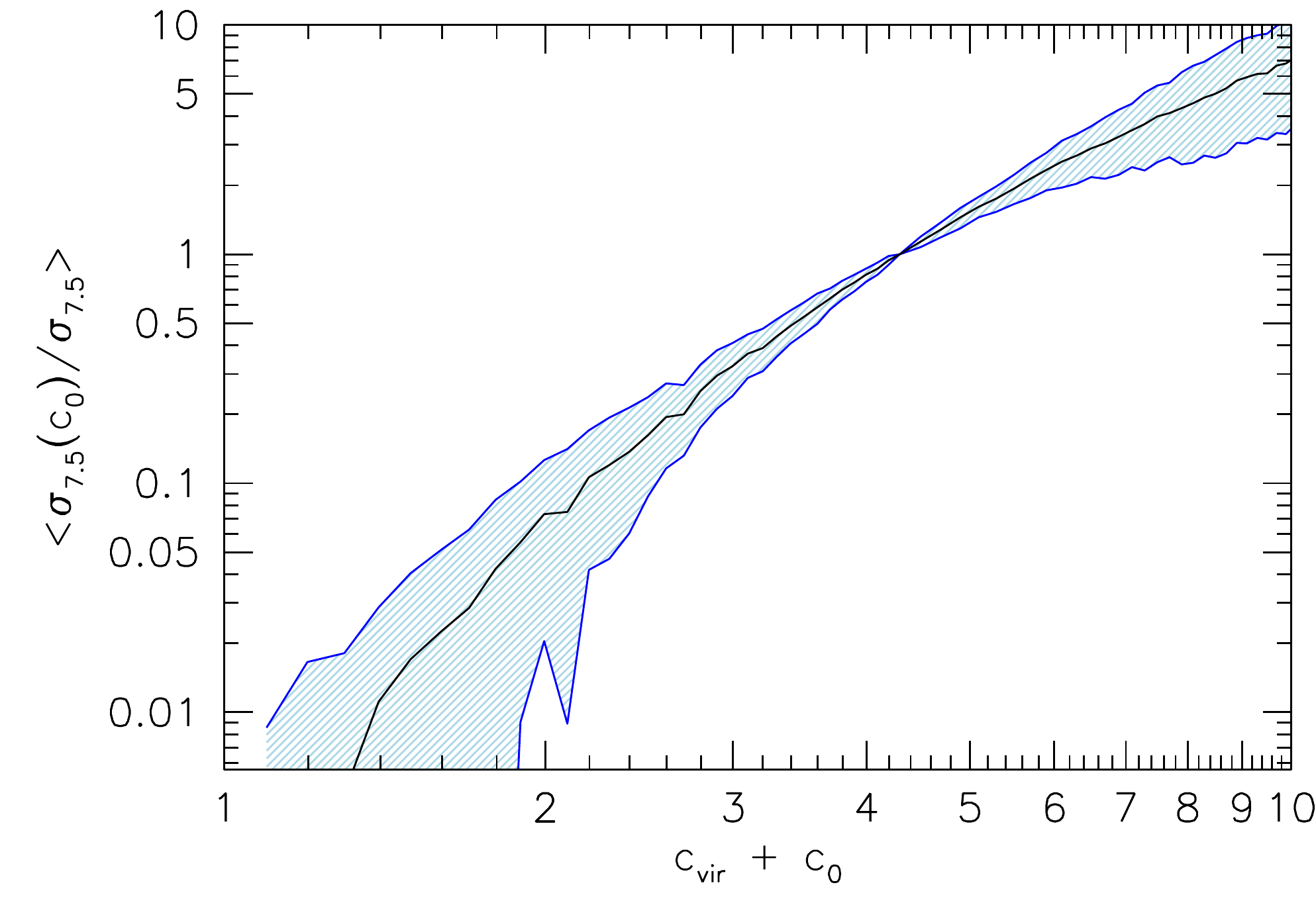}
\includegraphics[width=5.5cm]{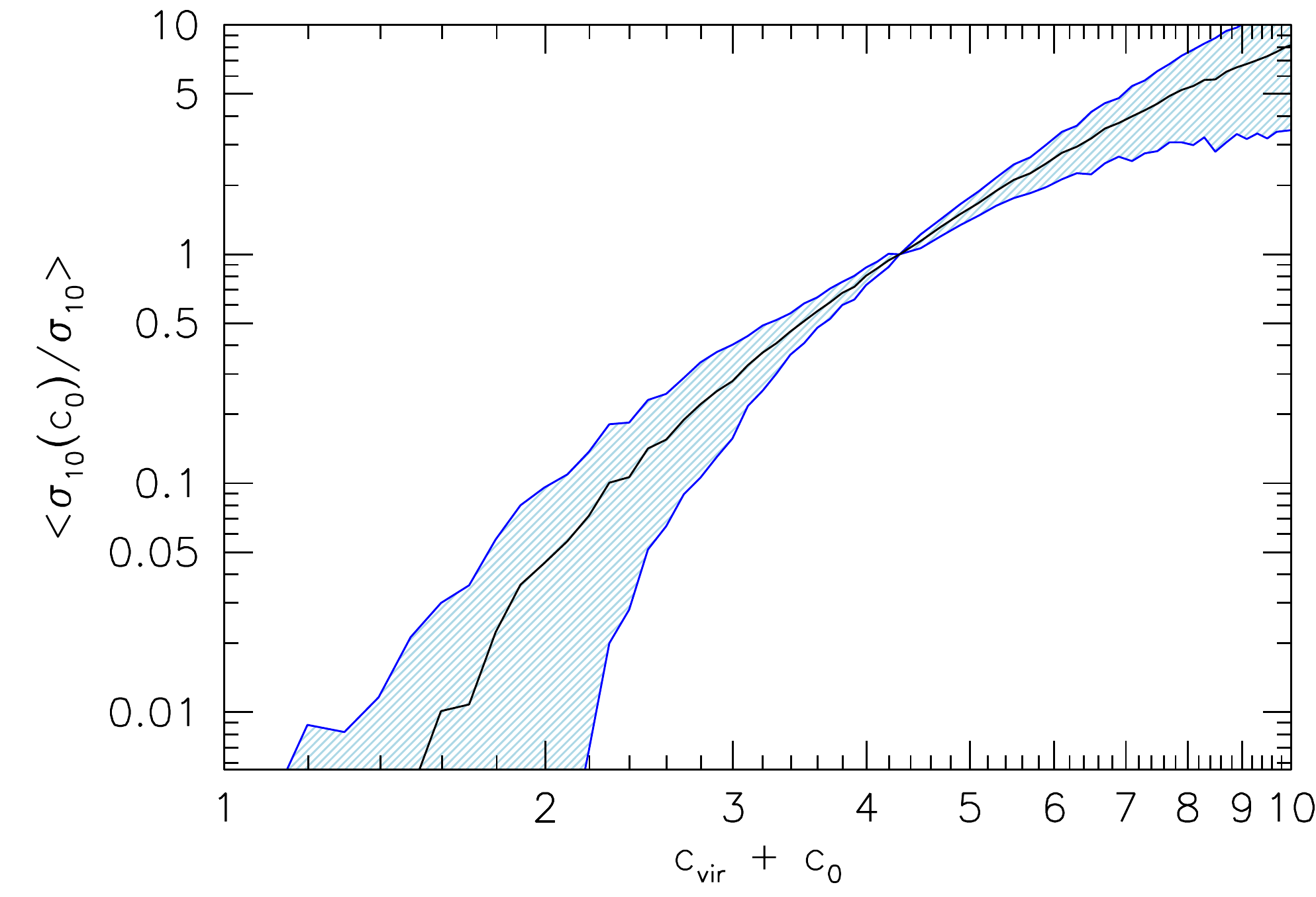}
\caption{The  median ratio of  the strong  lensing cross  section with
  respect to  the fiducial sample  as a function of  the normalization
  $c_{vir}+c_0$ in the  \citet{zhao09} mass concentration relation for
  three  different value  of  $l/w=5$,  $7.5$ and  $10$  in the  left,
  central  and  right  panels,  respectively.  The  shaded  region  is
  enclosed       by      the       lower      and       the      upper
  quartiles.\label{figmedianratiovsc}}
\end{figure*}

Figure  \ref{figmedianratiovsc}  shows  the  median  ratio  of  strong
lensing  cross sections  with our  fiducial  sample as  a function  of
$c_{vir}'=c_{vir}+c_0$   for  $l/w=5$,  $7.5$   and  $10$.   A  higher
normalization  tends to  increase strong  lensing. Going  from  a mean
value of  the concentration for cluster  size haloes from  $4$ to $8$,
the strong lensing cross section increases by about a factor of $4$.

\subsection{Subhalo abundance}
\citet{torri04}  have  shown  that  merger  events  and  substructures
increase the strong  lensing cross sections, mostly if  the latter are
located near the cluster centre.

Our analysis has so far  used cluster maps whose subhalo mass function
was obtained by sampling the distribution \ref{eqsubhalomassf} down to
$10^{10}M_{\odot}/h$. To statistically  analyse how the strong lensing
signal  depends on the  minimum subhalo  mass in  our models,  we have
generated a sample  of triaxial haloes with BCG,  sampling the subhalo
mass function  down to  a variable minimum  mass $m_{min}$.  The total
halo mass $M_{vir}=10^{15}M_{\odot}/h$ is kept fixed.

\begin{figure*}
\includegraphics[width=7.5cm]{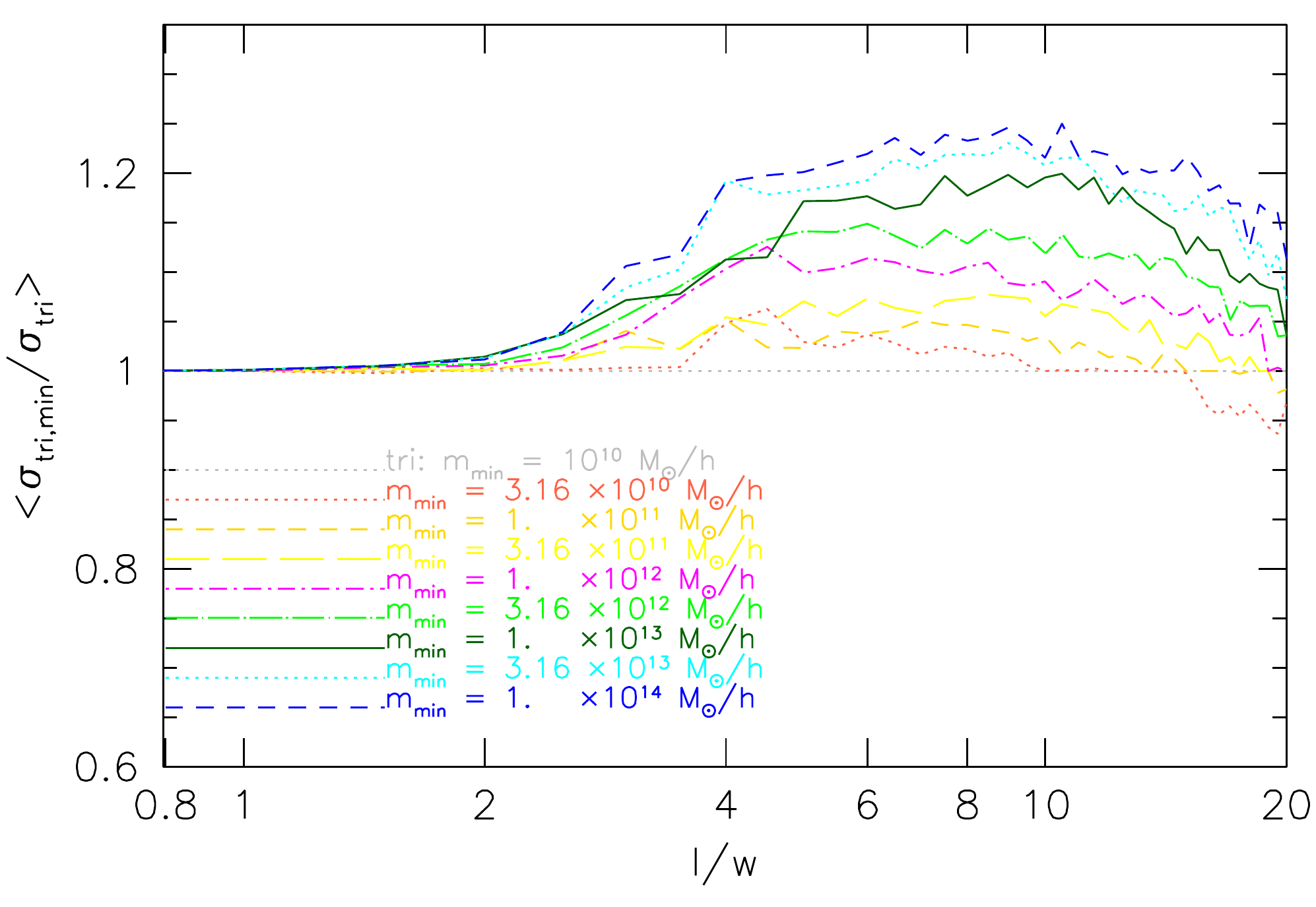}
\includegraphics[width=7.5cm]{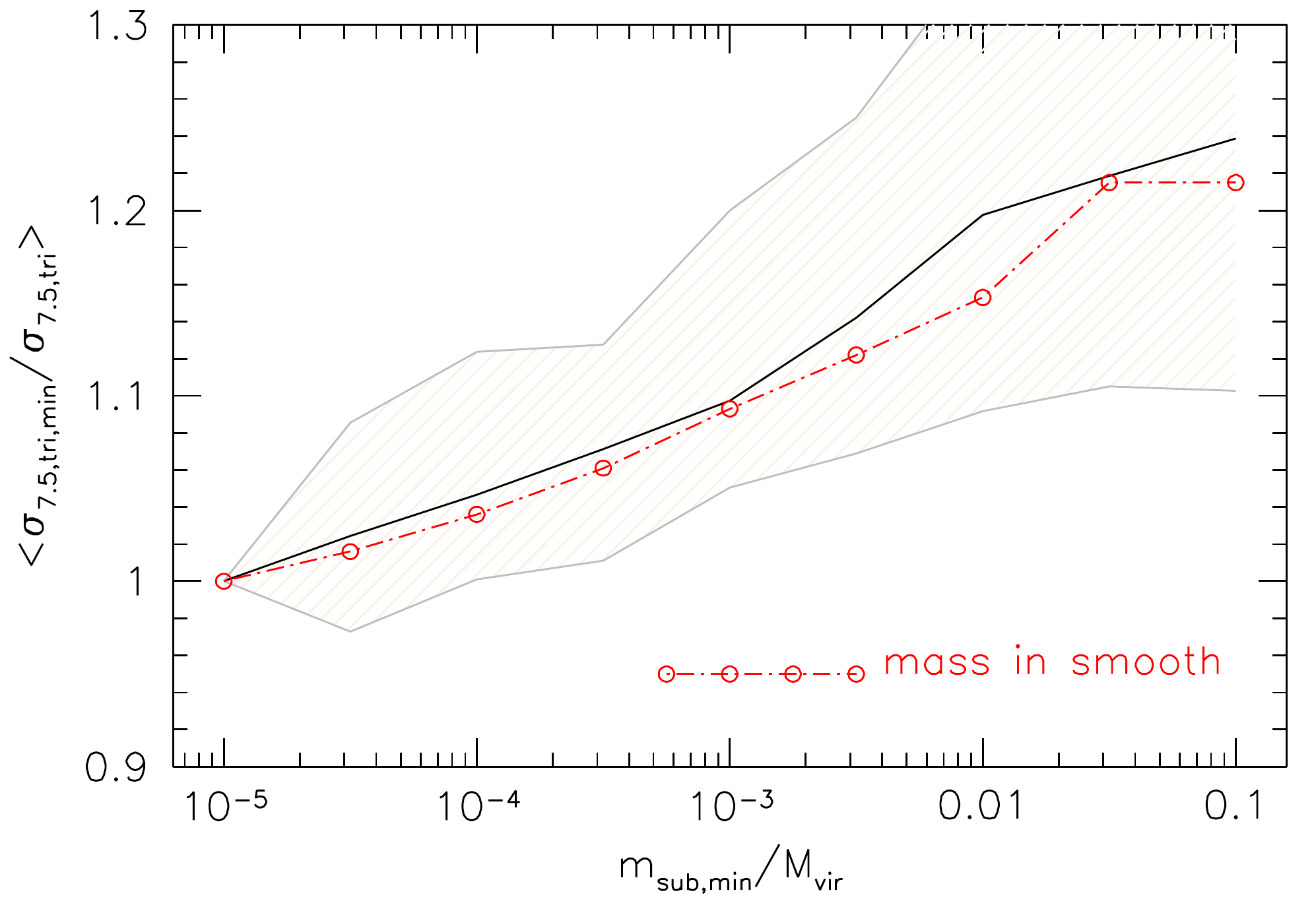}
\caption{The  dependence of the  strong lensing  cross section  on the
  minimum  subhalo  mass.  Left  panel: median  strong  lensing  cross
  section scaled by the fiducial triaxial simulation with subhalo mass
  resolution $10^{10}M_{\odot}/h$. Right  panel: median rescaled cross
  section at $l/w=7.5$ as a  function of the minimum subhalo mass. The
  median rescaled  strong lensing cross sections for  $l/w=5$ and $10$
  as a function of $m_{min}/M_{vir}$ do not fall far away from the one
  estimated for $l/w=7.5$. The data points connected with a dot-dashed
  line show the median of the host halo smooth mass component rescaled
  with respect to  the smooth mass component when  the minimum subhalo
  mass is  $10^{10}M_{\odot}/h$, as a function of  the minimum subhalo
  mass.  \label{figsubres}}
\end{figure*}

Figure \ref{figsubres}  shows the strong lensing signal  as a function
of the minimum subhalo mass. The  left panel shows the median ratio of
runs with  different resolutions compared to the  fiducial case, whose
subhalo  mass is  $\ge10^{10}M_{\odot}/h$. The  strong  lensing signal
increases    when   the    minimum   subhalo    mass    changes   from
$10^{10}M_{\odot}/h$ to  $10^{14}M_{\odot}/h$ because the  smooth halo
mass  tends to  increase for  larger value  of $m_{min}$,  raising the
projected  mass density distribution  near the  host halo  centre. The
right  panel shows the  ratio $\langle\sigma_{min}/\sigma  \rangle$ at
$l/w=7.5$ as a function of the minimum subhalo mass expressed in terms
of the total mass. Again,  the strong lensing signal increases because
the smooth  component contains  more mass. This  is shown in  the same
figure,  where the  data points  show the  median of  the  smooth mass
component in the host halo as  a function of the minimum subhalo mass,
rescaled with  respect to  the smooth mass  component, with  a minimum
subhalo mass of $10^{10}M_{\odot}/h$.

Notice that  the strong lensing  cross section depends on  the minimum
subhalo mass.  A decrease of the latter  by a factor of  $10$ tends to
increase   $\sigma_{7.5}$   by   $5\%$   because   the   smooth   mass
increases. However,  the impact of  substructures can be  different in
different clusters  depending of  the particular configuration  of the
lens

\subsection{Generalized NFW density profile}
One of the main potential problems  of the CDM model regards the inner
slope  of halo  density profiles.  Studying a  sample of  six strongly
lensing   clusters,  \citet{sand04}   concluded   that,  at   $68\,\%$
confidence, the inner slope is consistent with $\beta\approx 0.52$ and
that this is inconsistent with $\beta=1$ at the $99\,\%$ level. Recent
analyses  by  \citet{newman09,newman11}  of  Abell 611  and  383  have
confirmed a  flat central dark  matter density profile. How  much does
the inner slope bias the strong lensing signal?

To  answer this question  we have  generated a  sample of  haloes with
different $\beta$.  We take adiabatic  contraction in the  center into
account by  means of Eq.~(\ref{eqadcontraction}),  where $m_i(r_i)$ is
estimated by integrating equation (\ref{eqgNFW}).

The generalized NFW density profile
\begin{equation}
  \rho_{gNFW}=\frac{\rho_s}{(r/r_s)^{\beta}(1+r/r_s)^{3-\beta}}\,,
\label{eqgNFW}
\end{equation}
is  taken  to  have  an  arbitrary  inner  slope  $\beta$.  We  define
$r_{-2}\equiv      (2-\beta)r_s$       and      the      concentration
$c_{-2}=c_{vir}/(2-\beta)$ at the radius  where the density profile is
isothermal.

\begin{figure}
\includegraphics[width=\hsize]{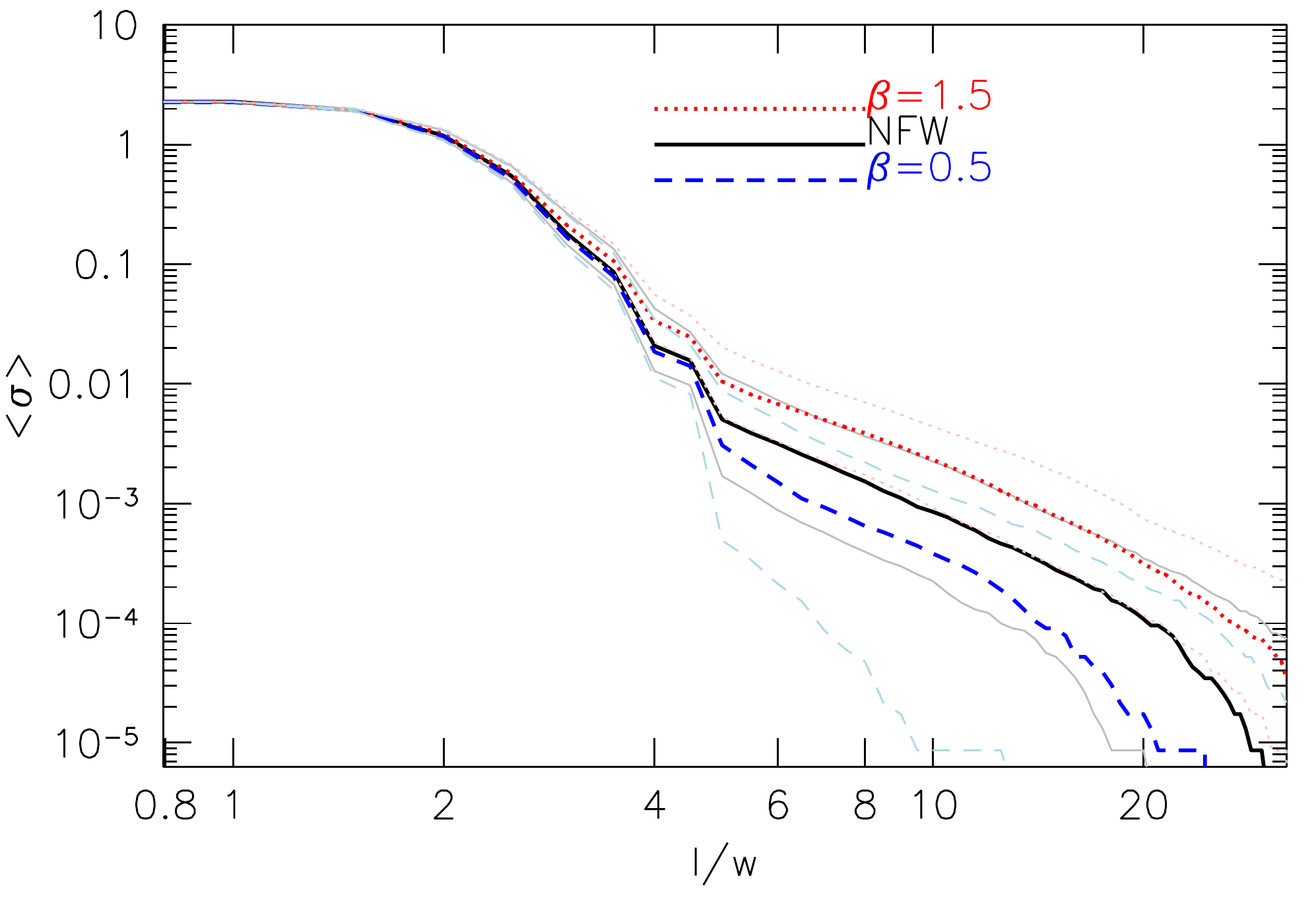}
\caption{The  median value  of the  strong lensing  cross  section for
  three different values of the  inner slope of the density profile of
  the host. Dotted, solid and dashed  line refer to a sample of haloes
  with $\beta=1.5$, $1$ and $0.5$, respectively. The thin solid curves
  show the quartiles of the NFW sample.\label{figbeta1}}
\end{figure}

Figure \ref{figbeta1}  shows the  median strong lensing  cross section
for three samples  of cluster sized haloes. $NFW$  labels our fiducial
sample  with  $\beta=1$,  while  $\beta=1.5$ and  $0.5$  illustrate  a
steeper and a shallower profile.

\begin{figure*}
\includegraphics[width=5.5cm]{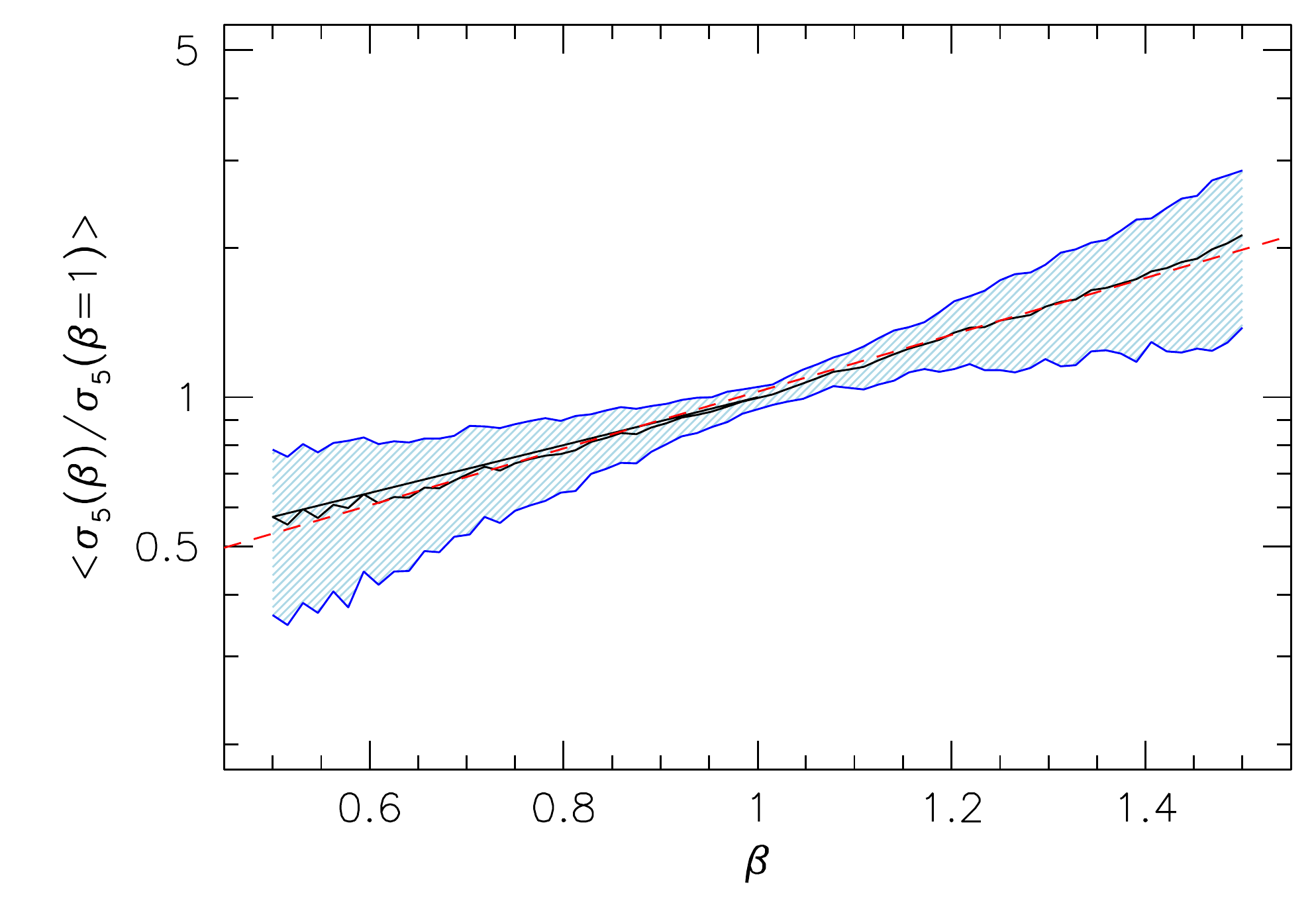}
\includegraphics[width=5.5cm]{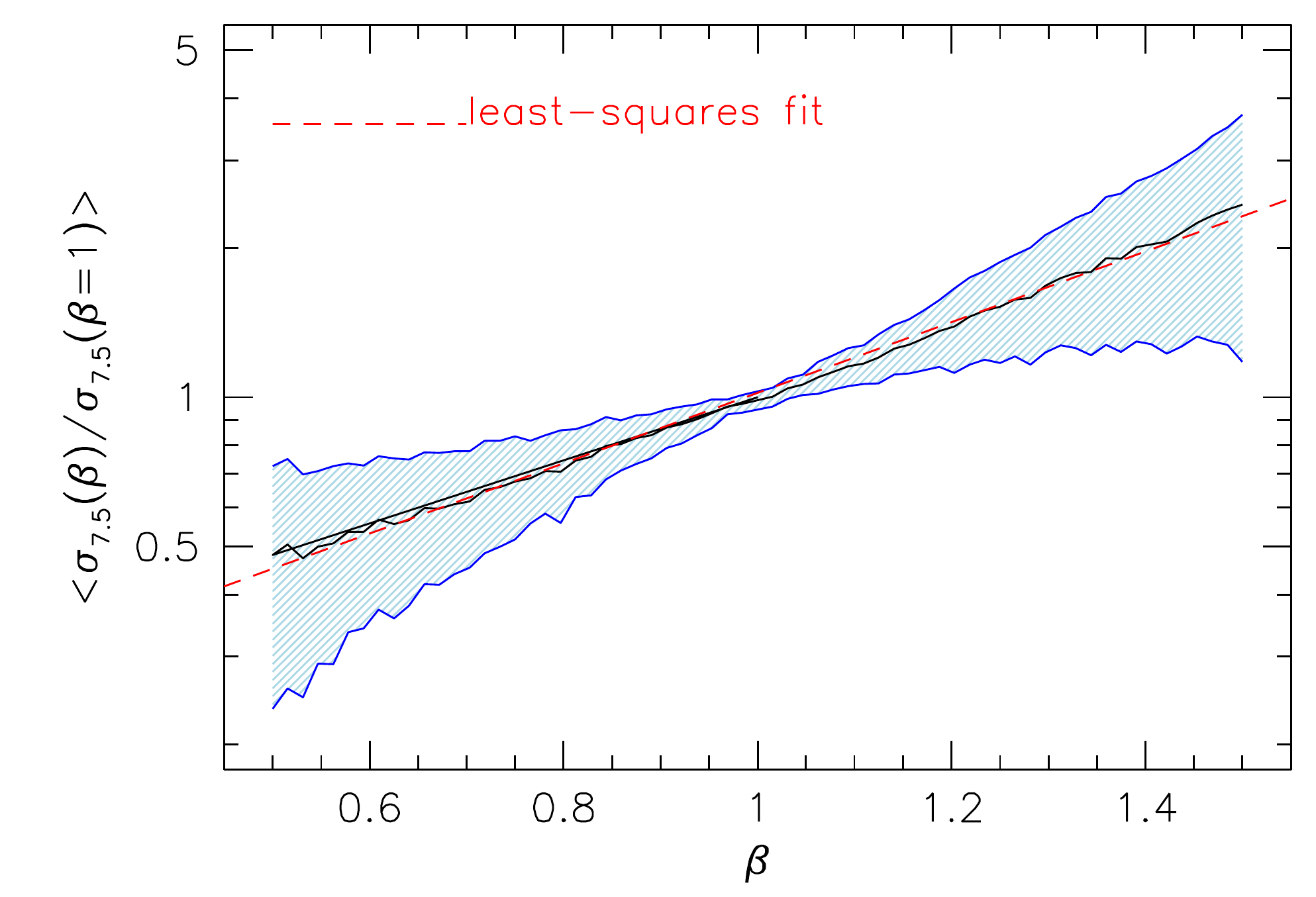}
\includegraphics[width=5.5cm]{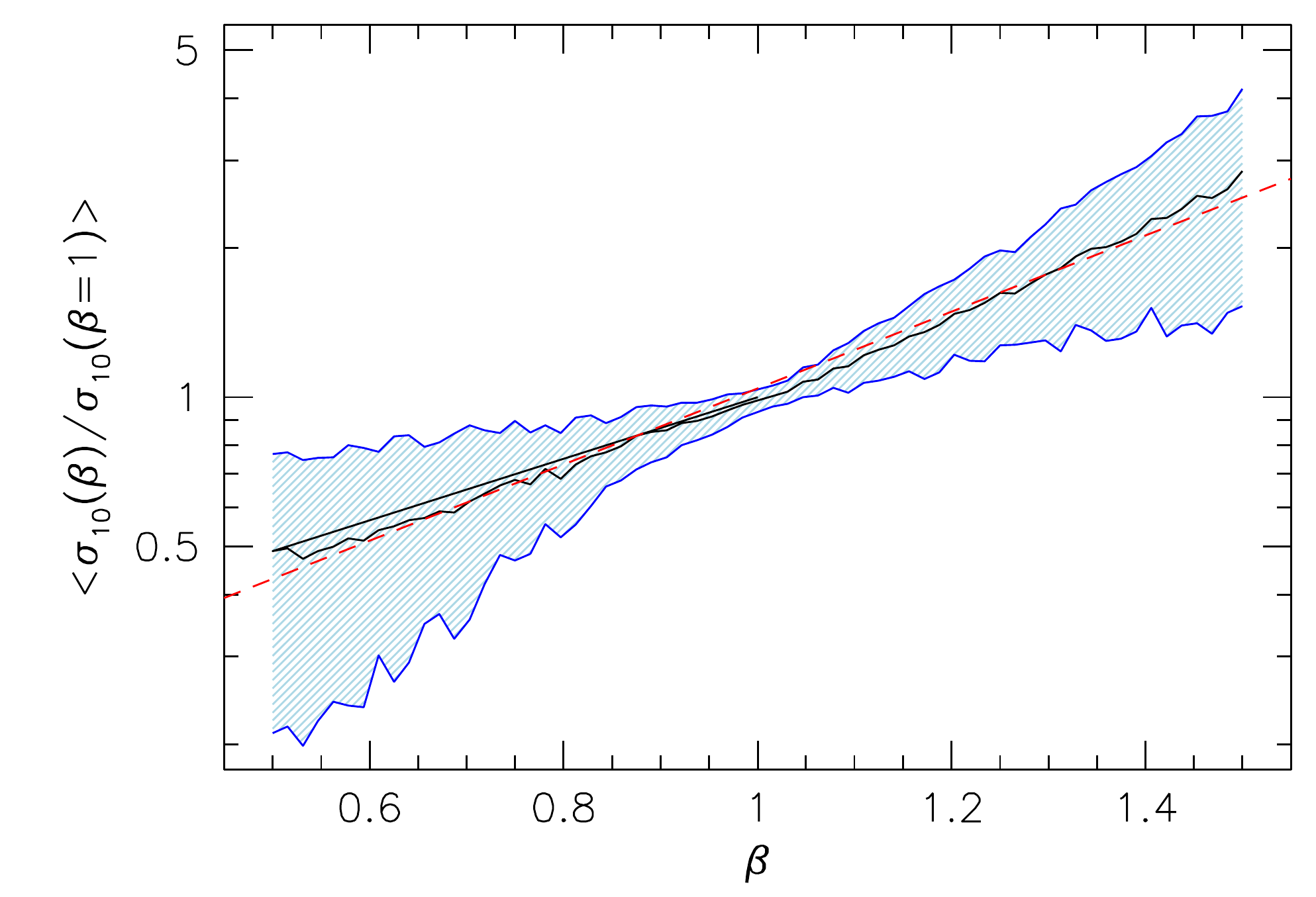}
\caption{The  median ratio  of  the strong  lensing  cross section  as
  function  the inner  slope  of  the main  halo  density profile  for
  $l/w=5$, $7.5$  and $10$.  In each panel  the shaded  region enclose
  $25-50\%$ of the  data and the dashed line  the least-squares fit to
  the correlation.\label{figbeta2}}
\end{figure*}

The three panels in  Fig.~\ref{figbeta2} show the relation between the
median ratio of the strong lensing cross section at $l/w=5$, $7.5$ and
$10$ compared  to our fiducial sample  as function of  the inner slope
$\beta$. In  each panel the shaded  region encloses the  lower and the
upper quartiles, while the dashed  line shows the least-squares fit to
the data,
\begin{equation}
  \log\left(\langle\sigma(\beta)/\sigma_{NFW}\rangle\right) = a \cdot\beta + b\,,
\end{equation}
for which we find
\begin{eqnarray}
  a=0.572 \pm 0.005,\; b=-0.560\pm 0.005 \quad &\sigma_{5}& \nonumber \\
  a=0.710 \pm 0.005,\; b=-0.701\pm 0.006 \quad &\sigma_{7.5}& \nonumber \\
  a=0.768 \pm 0.009,\; b=-0.750\pm 0.009 \quad &\sigma_{10}&. \nonumber 
\end{eqnarray}
A  shallower (steeper) inner  slope tends  to decrease  (increase) the
strong lensing  cross section with $l/w>5$ by  about $50\%$ ($100\%$).
Consindering triaxial haloes with $\beta=1$ and $\beta=1.5$, analogous
values for the  strong lensing cross section for  $l/w=7$ and $l/w=10$
and sources with $z_s<1.25$ have been found by \citet{oguri03}.

\section{Conclusions}
\label{disccon}
We have  presented a  new algorithm to  study the lensing  signal from
cluster  sized haloes. The  ingredients that  we use  to build  up our
triaxial and  substructured models take  into account the  most recent
results  from  numerical  simulations  of  structure  formation.   The
possibility of turning  these ingredients on and off  in our algorithm
allows us to quantify the  importance of all halo properties for their
strong  lensing efficiency.  Starting from  the halo  deflection angle
maps, we estimate strong lensing cross sections by ray-tracing.

We can summarize our main results as follows:
\begin{itemize}
\item Different structural halo properties affect strong lensing cross
  sections in different ways;
\item Averaging over a sample  of cluster-sized haloes, we find that a
  central galaxy  and triaxiality tend  to enhance the  strong lensing
  signal by $20\%$ and $50-70\%$, respectively;
\item the  strong lensing  cross section monotonically  increases with
  the  host  halo concentration;  in  a fixed  mass  bin,  it is  well
  characterized by log-normal distribution;
\item  an  increase  (decrease)  of  the  normalization  of  the  mass
  concentration relation  by a factor $c_0=2$  increases (reduces) the
  strong lensing cross section by a factor $5$ ($0.03$);
\item increasing the  minimum subhalo mass by a  factor of $10$ weakly
  increases $\sigma_{7.5}$ by about $3\%$;
\item the change in the inner slope of the density profile of the host
  halo  is   linearly  related  to   $y=\log(\sigma_{l/w})$;  profiles
  shallower than NFW are weaker strong lenses;
\end{itemize}
These  results may help  to understand  future observational  data and
predictions for upcoming wide field surveys \citep{refregier10}.

\section*{Anknowledgments}
Many  thanks to  the anonymous  referee for  the useful  comments that
helped us to improve the  presentation of the paper.  Thanks to Cosimo
Fedeli for  an useful  discussion during a  hot day in  Bologna.  This
work was  supported by the  EU-RTN "DUEL", ASI  contracts: I/009/10/0,
EUCLID-IC  fase A/B1  and  PRIN-INAF 2009  and  by the  DAAD and  CRUI
through   their   Vigoni   programme.    LM   acknowledges   financial
contributions from contracts  ASIINAF I/023/05/0, ASI-INAF I/088/06/0,
ASI   I/016/07/0   "COFIS",   ASI   "Euclid-DUNE"   I/064/08/0,ASI-Uni
Bologna-Astronomy  Dept. Euclid-NIS  I/039/10/0, and  PRIN  MIUR "Dark
energy and cosmology with large galaxy surveys"

\appendix
\section{Scatter in the strong lensing cross section}
\label{appcrel}
Haloes form hierarchically by gravitational instability of dark matter
density  fluctuations. Small  systems  form first  and  merge to  form
larger objects. The assembly  history depends on the environment where
a halo grows. Virialized structures  with the same final mass may have
experienced different growth histories \citep{gao05a}. Haloes with the
same  mass can  thus have  different concentrations  and  subhalo mass
functions.  Numerical simulations  have shown  that  the concentration
distribution is well fit by a log-normal distribution
\begin{equation}
  p(c|M_{vir}) = \dfrac{1}{\sqrt{2 \pi \sigma_{\ln c}^2}}
  \exp\left[ - \dfrac{\left(\ln c - \ln c_{vir}\right)^2}{2\sigma_{\ln c}^2} \right]\,.
\end{equation}

\begin{figure}
\includegraphics[width=\hsize]{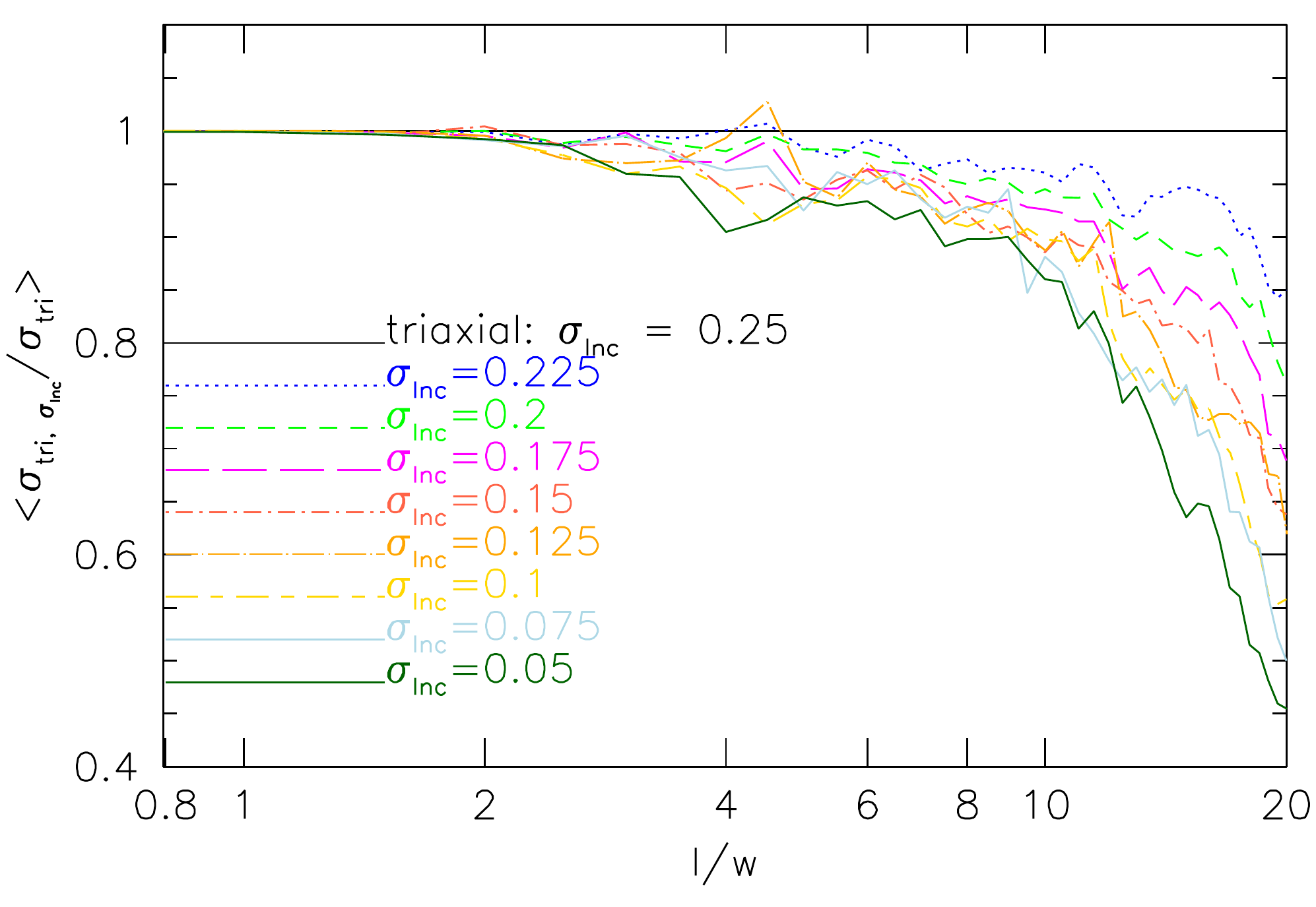}
\caption{The median ratio of  the strong lensing cross section between
  sample of  haloes generated assuming a scatter  $\sigma_{\ln c}$ and
  our   fiducial   one  --   where   we   have  assumed   $\sigma_{\ln
    c}=0.25$.  Different line types  refer to  different value  of the
  variance  in   the  log-normal  scatter  from  which   we  draw  the
  concentration  to  assign  to  each  host  halo  in  the  respective
  sample.  \label{figsigcmedian}}
\end{figure}

We assume $\sigma_{\ln  c}=0.25$ in our algorithm. In  order to relate
the  concentration scatter with  the scatter  of strong  lensing cross
sections at  fixed host  halo mass, we  have generated  different halo
samples with  different $\sigma_{\ln c}$  from which we  randomly draw
the  host  halo concentration.  Figure  \ref{figsigcmedian} shows  the
median  ratio of  the strong  lensing cross  sections compared  to our
fiducial sample for different standard deviations $\sigma_{\ln c}$. We
find that smaller standard deviations reduce the median strong lensing
signal.

\begin{figure}
\includegraphics[width=\hsize]{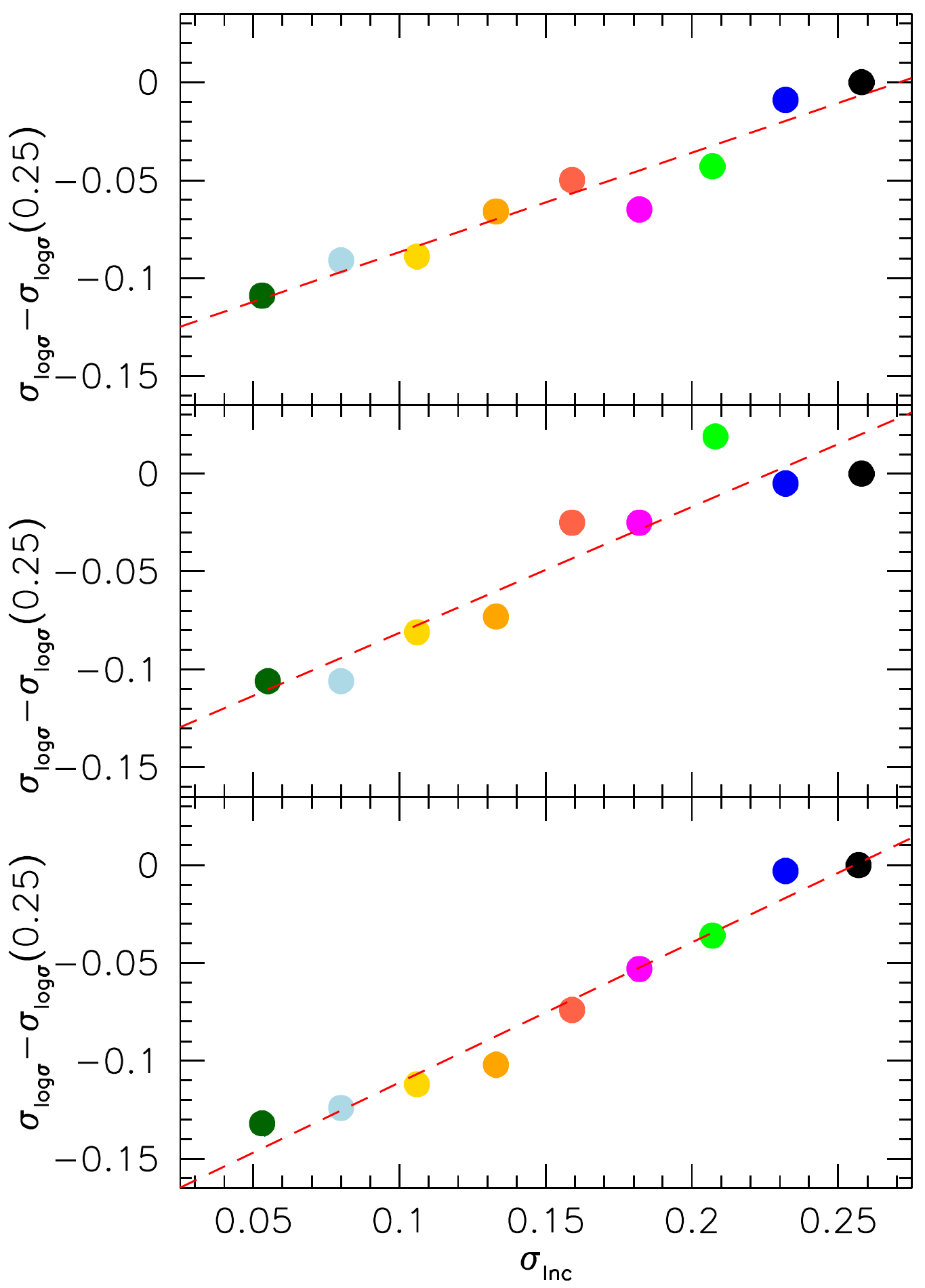}
\caption{The  correlation between  the scatter  in the  strong lensing
  cross section at  three different value of $l/w$  and the scatter in
  concentration         assumed        to         generate        halo
  samples.  \label{figcorrsigma}}
\end{figure}

For  each  halo  sample,  we  have estimated  the  standard  deviation
$\sigma_{\log \sigma}$  for length-to-width ratios  $l/w=5$, $7.5$ and
$10$.  Figure  \ref{figcorrsigma}   shows  the  relation  between  the
standard  deviation  in the  strong  lensing  signal  for haloes  with
$M_{vir}=10^{15}M_{\odot}/h$, $z_l=0.25$  and $z_s=2$ and  the scatter
in  concentration,  rescaled  to  the  scatter  for  the  sample  with
$\sigma_{\ln c}=0.25$. For the chosen $l/w$ ratios, the correlation is
well  fit  by a  straight  line.  Higher  standard deviations  in  the
concentration scatter cause larger  strong lensing signals. The dashed
lines in all panels show the least-squares fit to the data points,
\begin{equation}
\sigma_{\log \sigma} = a \times \sigma_{\ln c} + b\,,
\end{equation}
where we obtain the slopes
\begin{eqnarray}
  a=0.714 \pm 0.049, \quad \mathrm{for}\quad &\sigma_{5}& \nonumber \\
  a=0.640 \pm 0.091, \quad \mathrm{for}\quad &\sigma_{7.5}& \nonumber \\
  a=0.506 \pm 0.053, \quad \mathrm{for}\quad &\sigma_{10}&. \nonumber 
\end{eqnarray}

\bibliographystyle{mn2e}
\bibliography{geuclidSL}

\begin{thebibliography}{}

\bibitem[\protect\citeauthoryear{{Bartelmann}}{{Bartelmann}}{1996}]{bartelmann96}
{Bartelmann} M.,  1996, \aap, 313, 697

\bibitem[\protect\citeauthoryear{{Bartelmann}, {Huss}, {Colberg}, {Jenkins} \&
  {Pearce}}{{Bartelmann} et~al.}{1998}]{bartelmann98}
{Bartelmann} M.,  {Huss} A.,  {Colberg} J.~M.,  {Jenkins} A.,    {Pearce}
  F.~R.,  1998, \aap, 330, 1

\bibitem[\protect\citeauthoryear{{Bartelmann}, {Meneghetti}, {Perrotta},
  {Baccigalupi} \& {Moscardini}}{{Bartelmann} et~al.}{2003}]{bartelmann03}
{Bartelmann} M.,  {Meneghetti} M.,  {Perrotta} F.,  {Baccigalupi} C.,
  {Moscardini} L.,  2003, \aap, 409, 449

\bibitem[\protect\citeauthoryear{{Bartelmann} \& {Schneider}}{{Bartelmann} \&
  {Schneider}}{2001}]{bartelmann01}
{Bartelmann} M.,  {Schneider} P.,  2001, Physics Reports, 340, 291

\bibitem[\protect\citeauthoryear{{Blumenthal}, {Faber}, {Flores} \&
  {Primack}}{{Blumenthal} et~al.}{1986}]{blumenthal86}
{Blumenthal} G.~R.,  {Faber} S.~M.,  {Flores} R.,    {Primack} J.~R.,  1986,
  \apj, 301, 27

\bibitem[\protect\citeauthoryear{{Bond}, {Cole}, {Efstathiou} \&
  {Kaiser}}{{Bond} et~al.}{1991}]{bond91}
{Bond} J.~R.,  {Cole} S.,  {Efstathiou} G.,    {Kaiser} N.,  1991, \apj, 379,
  440

\bibitem[\protect\citeauthoryear{{Bryan} \& {Norman}}{{Bryan} \&
  {Norman}}{1998}]{bryan98}
{Bryan} G.~L.,  {Norman} M.~L.,  1998, \apj, 495, 80

\bibitem[\protect\citeauthoryear{{Bullock}, {Kolatt}, {Sigad}, {Somerville},
  {Kravtsov}, {Klypin}, {Primack} \& {Dekel}}{{Bullock}
  et~al.}{2001}]{bullock01a}
{Bullock} J.~S.,  {Kolatt} T.~S.,  {Sigad} Y.,  {Somerville} R.~S.,  {Kravtsov}
  A.~V.,  {Klypin} A.~A.,  {Primack} J.~R.,    {Dekel} A.,  2001, \mnras, 321,
  559

\bibitem[\protect\citeauthoryear{{Choi}, {Weinberg} \& {Katz}}{{Choi}
  et~al.}{2007}]{choi07}
{Choi} J.-H.,  {Weinberg} M.~D.,    {Katz} N.,  2007, \mnras, 381, 987

\bibitem[\protect\citeauthoryear{{Croton}, {Springel}, {White}, {De Lucia},
  {Frenk}, {Gao}, {Jenkins}, {Kauffmann}, {Navarro} \& {Yoshida}}{{Croton}
  et~al.}{2006}]{croton06}
{Croton} D.~J.,  {Springel} V.,  {White} S.~D.~M.,  {De Lucia} G.,  {Frenk}
  C.~S.,  {Gao} L.,  {Jenkins} A.,  {Kauffmann} G.,  {Navarro} J.~F.,
  {Yoshida} N.,  2006, \mnras, 365, 11

\bibitem[\protect\citeauthoryear{{De Lucia}, {Kauffmann}, {Springel}, {White},
  {Lanzoni}, {Stoehr}, {Tormen} \& {Yoshida}}{{De Lucia}
  et~al.}{2004}]{delucia04}
{De Lucia} G.,  {Kauffmann} G.,  {Springel} V.,  {White} S.~D.~M.,  {Lanzoni}
  B.,  {Stoehr} F.,  {Tormen} G.,    {Yoshida} N.,  2004, \mnras, 348, 333

\bibitem[\protect\citeauthoryear{{Dolag}, {Bartelmann}, {Perrotta},
  {Baccigalupi}, {Moscardini}, {Meneghetti} \& {Tormen}}{{Dolag}
  et~al.}{2004}]{dolag04}
{Dolag} K.,  {Bartelmann} M.,  {Perrotta} F.,  {Baccigalupi} C.,  {Moscardini}
  L.,  {Meneghetti} M.,    {Tormen} G.,  2004, \aap, 416, 853

\bibitem[\protect\citeauthoryear{{Duffy}, {Schaye}, {Kay} \& {Dalla
  Vecchia}}{{Duffy} et~al.}{2008}]{duffy08}
{Duffy} A.~R.,  {Schaye} J.,  {Kay} S.~T.,    {Dalla Vecchia} C.,  2008,
  \mnras, 390, L64

\bibitem[\protect\citeauthoryear{{Eke}, {Cole} \& {Frenk}}{{Eke}
  et~al.}{1996}]{eke96}
{Eke} V.~R.,  {Cole} S.,    {Frenk} C.~S.,  1996, \mnras, 282, 263

\bibitem[\protect\citeauthoryear{{Fasano}, {Bettoni}, {Ascaso}, {Tormen},
  {Poggianti}, {Valentinuzzi}, {D'Onofrio}, {Fritz}, {Moretti}, {Omizzolo},
  {Cava}, {Moles}, {Dressler}, {Couch}, {Kj{\ae}rgaard} \& {Varela}}{{Fasano}
  et~al.}{2010}]{fasano10}
{Fasano} G.,  {Bettoni} D.,  {Ascaso} B.,  {Tormen} G.,  {Poggianti} B.~M.,
  {Valentinuzzi} T.,  {D'Onofrio} M.,  {Fritz} J.,  {Moretti} A.,  {Omizzolo}
  A.,  {Cava} A.,  {Moles} M.,  {Dressler} A.,  {Couch} W.~J.,  {Kj{\ae}rgaard}
  P.,    {Varela} J.,  2010, \mnras, 404, 1490

\bibitem[\protect\citeauthoryear{{Fedeli}, {Meneghetti}, {Bartelmann}, {Dolag}
  \& {Moscardini}}{{Fedeli} et~al.}{2006}]{fedeli06}
{Fedeli} C.,  {Meneghetti} M.,  {Bartelmann} M.,  {Dolag} K.,    {Moscardini}
  L.,  2006, \aap, 447, 419

\bibitem[\protect\citeauthoryear{{Fedeli}, {Meneghetti}, {Gottl{\"o}ber} \&
  {Yepes}}{{Fedeli} et~al.}{2010}]{fedeli10}
{Fedeli} C.,  {Meneghetti} M.,  {Gottl{\"o}ber} S.,    {Yepes} G.,  2010, \aap,
  519, A91+

\bibitem[\protect\citeauthoryear{{Gao}, {Navarro}, {Cole}, {Frenk}, {White},
  {Springel}, {Jenkins} \& {Neto}}{{Gao} et~al.}{2008}]{gao08}
{Gao} L.,  {Navarro} J.~F.,  {Cole} S.,  {Frenk} C.~S.,  {White} S.~D.~M.,
  {Springel} V.,  {Jenkins} A.,    {Neto} A.~F.,  2008, \mnras, 387, 536

\bibitem[\protect\citeauthoryear{{Gao}, {Springel} \& {White}}{{Gao}
  et~al.}{2005}]{gao05a}
{Gao} L.,  {Springel} V.,    {White} S.~D.~M.,  2005, \mnras, 363, L66

\bibitem[\protect\citeauthoryear{{Gao}, {White}, {Jenkins}, {Stoehr} \&
  {Springel}}{{Gao} et~al.}{2004}]{gao04}
{Gao} L.,  {White} S.~D.~M.,  {Jenkins} A.,  {Stoehr} F.,    {Springel} V.,
  2004, \mnras, 355, 819

\bibitem[\protect\citeauthoryear{{Giocoli}, {Bartelmann}, {Sheth} \&
  {Cacciato}}{{Giocoli} et~al.}{2010}]{giocoli10b}
{Giocoli} C.,  {Bartelmann} M.,  {Sheth} R.~K.,    {Cacciato} M.,  2010,
  \mnras, 408, 300

\bibitem[\protect\citeauthoryear{{Giocoli}, {Moreno}, {Sheth} \&
  {Tormen}}{{Giocoli} et~al.}{2007}]{giocoli07a}
{Giocoli} C.,  {Moreno} J.,  {Sheth} R.~K.,    {Tormen} G.,  2007, \mnras, 376,
  977

\bibitem[\protect\citeauthoryear{{Giocoli}, {Tormen}, {Sheth} \& {van den
  Bosch}}{{Giocoli} et~al.}{2010a}]{giocoli10}
{Giocoli} C.,  {Tormen} G.,  {Sheth} R.~K.,    {van den Bosch} F.~C.,  2010a,
  \mnras, 404, 502

\bibitem[\protect\citeauthoryear{{Giocoli}, {Tormen} \& {van den
  Bosch}}{{Giocoli} et~al.}{2008}]{giocoli08b}
{Giocoli} C.,  {Tormen} G.,    {van den Bosch} F.~C.,  2008, \mnras, 386, 2135

\bibitem[\protect\citeauthoryear{{Hayashi}, {Navarro}, {Taylor}, {Stadel} \&
  {Quinn}}{{Hayashi} et~al.}{2003}]{hayashi03}
{Hayashi} E.,  {Navarro} J.~F.,  {Taylor} J.~E.,  {Stadel} J.,    {Quinn} T.,
  2003, \apj, 584, 541

\bibitem[\protect\citeauthoryear{{Hernquist}}{{Hernquist}}{1990}]{hernquist90}
{Hernquist} L.,  1990, \apj, 356, 359

\bibitem[\protect\citeauthoryear{{Jing}}{{Jing}}{2000}]{jing00}
{Jing} Y.~P.,  2000, \apj, 535, 30

\bibitem[\protect\citeauthoryear{{Jing} \& {Suto}}{{Jing} \&
  {Suto}}{2002}]{jing02}
{Jing} Y.~P.,  {Suto} Y.,  2002, \apj, 574, 538

\bibitem[\protect\citeauthoryear{{Kazantzidis}, {Kravtsov}, {Zentner},
  {Allgood}, {Nagai} \& {Moore}}{{Kazantzidis} et~al.}{2004}]{kazantzidis04}
{Kazantzidis} S.,  {Kravtsov} A.~V.,  {Zentner} A.~R.,  {Allgood} B.,  {Nagai}
  D.,    {Moore} B.,  2004, \apjl, 611, L73

\bibitem[\protect\citeauthoryear{{Keeton}}{{Keeton}}{2001}]{keeton01}
{Keeton} C.~R.,  2001, \apj, 561, 46

\bibitem[\protect\citeauthoryear{{Keeton}}{{Keeton}}{2003}]{keeton03}
{Keeton} C.~R.,  2003, \apj, 584, 664

\bibitem[\protect\citeauthoryear{{Lacey} \& {Cole}}{{Lacey} \&
  {Cole}}{1993}]{lacey93}
{Lacey} C.,  {Cole} S.,  1993, \mnras, 262, 627

\bibitem[\protect\citeauthoryear{{Mandelbaum}, {van de Ven} \&
  {Keeton}}{{Mandelbaum} et~al.}{2009}]{mandelbaum09}
{Mandelbaum} R.,  {van de Ven} G.,    {Keeton} C.~R.,  2009, \mnras, 398, 635

\bibitem[\protect\citeauthoryear{{Meneghetti}, {Argazzi}, {Pace}, {Moscardini},
  {Dolag}, {Bartelmann}, {Li} \& {Oguri}}{{Meneghetti}
  et~al.}{2007b}]{meneghetti07b}
{Meneghetti} M.,  {Argazzi} R.,  {Pace} F.,  {Moscardini} L.,  {Dolag} K.,
  {Bartelmann} M.,  {Li} G.,    {Oguri} M.,  2007b, \aap, 461, 25

\bibitem[\protect\citeauthoryear{{Meneghetti}, {Bartelmann}, {Dolag},
  {Moscardini}, {Perrotta}, {Baccigalupi} \& {Tormen}}{{Meneghetti}
  et~al.}{2005a}]{meneghetti05a}
{Meneghetti} M.,  {Bartelmann} M.,  {Dolag} K.,  {Moscardini} L.,  {Perrotta}
  F.,  {Baccigalupi} C.,    {Tormen} G.,  2005a, \aap, 442, 413

\bibitem[\protect\citeauthoryear{{Meneghetti}, {Bartelmann} \&
  {Moscardini}}{{Meneghetti} et~al.}{2003}]{meneghetti03}
{Meneghetti} M.,  {Bartelmann} M.,    {Moscardini} L.,  2003, \mnras, 346, 67

\bibitem[\protect\citeauthoryear{{Meneghetti}, {Fedeli}, {Pace},
  {Gottl{\"o}ber} \& {Yepes}}{{Meneghetti} et~al.}{2010a}]{meneghetti10a}
{Meneghetti} M.,  {Fedeli} C.,  {Pace} F.,  {Gottl{\"o}ber} S.,    {Yepes} G.,
  2010a, \aap, 519, A90+

\bibitem[\protect\citeauthoryear{{Meneghetti}, {Fedeli}, {Zitrin},
  {Bartelmann}, {Broadhurst}, {Gottl{\"o}ber}, {Moscardini} \&
  {Yepes}}{{Meneghetti} et~al.}{2011}]{meneghetti11}
{Meneghetti} M.,  {Fedeli} C.,  {Zitrin} A.,  {Bartelmann} M.,  {Broadhurst}
  T.,  {Gottl{\"o}ber} S.,  {Moscardini} L.,    {Yepes} G.,  2011, \aap, 530,
  A17+

\bibitem[\protect\citeauthoryear{{Meneghetti}, {Jain}, {Bartelmann} \&
  {Dolag}}{{Meneghetti} et~al.}{2005b}]{meneghetti05b}
{Meneghetti} M.,  {Jain} B.,  {Bartelmann} M.,    {Dolag} K.,  2005b, \mnras,
  362, 1301

\bibitem[\protect\citeauthoryear{{Meneghetti}, {Rasia}, {Merten}, {Bellagamba},
  {Ettori}, {Mazzotta}, {Dolag} \& {Marri}}{{Meneghetti}
  et~al.}{2010b}]{meneghetti10b}
{Meneghetti} M.,  {Rasia} E.,  {Merten} J.,  {Bellagamba} F.,  {Ettori} S.,
  {Mazzotta} P.,  {Dolag} K.,    {Marri} S.,  2010b, \aap, 514, A93+

\bibitem[\protect\citeauthoryear{{Metcalf} \& {Madau}}{{Metcalf} \&
  {Madau}}{2001}]{metcalf01}
{Metcalf} R.~B.,  {Madau} P.,  2001, \mnras, 563, 9

\bibitem[\protect\citeauthoryear{{Moore}, {Ghigna}, {Governato}, {Lake},
  {Quinn}, {Stadel} \& {Tozzi}}{{Moore} et~al.}{1999}]{moore99}
{Moore} B.,  {Ghigna} S.,  {Governato} F.,  {Lake} G.,  {Quinn} T.,  {Stadel}
  J.,    {Tozzi} P.,  1999, \apjl, 524, L19

\bibitem[\protect\citeauthoryear{{Navarro}, {Frenk} \& {White}}{{Navarro}
  et~al.}{1996}]{navarro96}
{Navarro} J.~F.,  {Frenk} C.~S.,    {White} S.~D.~M.,  1996, \apj, 462, 563

\bibitem[\protect\citeauthoryear{{Neto}, {Gao}, {Bett}, {Cole}, {Navarro},
  {Frenk}, {White}, {Springel} \& {Jenkins}}{{Neto} et~al.}{2007}]{neto07}
{Neto} A.~F.,  {Gao} L.,  {Bett} P.,  {Cole} S.,  {Navarro} J.~F.,  {Frenk}
  C.~S.,  {White} S.~D.~M.,  {Springel} V.,    {Jenkins} A.,  2007, \mnras,
  381, 1450

\bibitem[\protect\citeauthoryear{{Newman}, {Treu}, {Ellis} \& {Sand}}{{Newman}
  et~al.}{2011}]{newman11}
{Newman} A.~B.,  {Treu} T.,  {Ellis} R.~S.,    {Sand} D.~J.,  2011, \apjl, 728,
  L39+

\bibitem[\protect\citeauthoryear{{Newman}, {Treu}, {Ellis}, {Sand}, {Richard},
  {Marshall}, {Capak} \& {Miyazaki}}{{Newman} et~al.}{2009}]{newman09}
{Newman} A.~B.,  {Treu} T.,  {Ellis} R.~S.,  {Sand} D.~J.,  {Richard} J.,
  {Marshall} P.~J.,  {Capak} P.,    {Miyazaki} S.,  2009, \apj, 706, 1078

\bibitem[\protect\citeauthoryear{{Oguri}, {Hennawi}, {Gladders}, {Dahle},
  {Natarajan}, {Dalal}, {Koester}, {Sharon} \& {Bayliss}}{{Oguri}
  et~al.}{2009}]{oguri09}
{Oguri} M.,  {Hennawi} J.~F.,  {Gladders} M.~D.,  {Dahle} H.,  {Natarajan} P.,
  {Dalal} N.,  {Koester} B.~P.,  {Sharon} K.,    {Bayliss} M.,  2009, \apj,
  699, 1038

\bibitem[\protect\citeauthoryear{{Oguri}, {Lee} \& {Suto}}{{Oguri}
  et~al.}{2003}]{oguri03}
{Oguri} M.,  {Lee} J.,    {Suto} Y.,  2003, \apj, 599, 7

\bibitem[\protect\citeauthoryear{{Oguri}, {Takada}, {Umetsu} \&
  {Broadhurst}}{{Oguri} et~al.}{2005}]{oguri05}
{Oguri} M.,  {Takada} M.,  {Umetsu} K.,    {Broadhurst} T.,  2005, \apj, 632,
  841

\bibitem[\protect\citeauthoryear{{Press} \& {Schechter}}{{Press} \&
  {Schechter}}{1974}]{press74}
{Press} W.~H.,  {Schechter} P.,  1974, \apj, 187, 425

\bibitem[\protect\citeauthoryear{{Puchwein}, {Bartelmann}, {Dolag} \&
  {Meneghetti}}{{Puchwein} et~al.}{2005}]{puchwein05}
{Puchwein} E.,  {Bartelmann} M.,  {Dolag} K.,    {Meneghetti} M.,  2005, \aap,
  442, 405

\bibitem[\protect\citeauthoryear{{Refregier}, {Amara}, {Kitching}, {Rassat},
  {Scaramella}, {Weller} \& {Euclid Imaging Consortium}}{{Refregier}
  et~al.}{2010}]{refregier10}
{Refregier} A.,  {Amara} A.,  {Kitching} T.~D.,  {Rassat} A.,  {Scaramella} R.,
   {Weller} J.,    {Euclid Imaging Consortium} f.~t.,  2010, ArXiv e-prints

\bibitem[\protect\citeauthoryear{{Richard}, {Smith}, {Kneib}, {Ellis},
  {Sanderson}, {Pei}, {Targett}, {Sand}, {Swinbank}, {Dannerbauer}, {Mazzotta},
  {Limousin}, {Egami}, {Jullo}, {Hamilton-Morris} \& {Moran}}{{Richard}
  et~al.}{2010}]{richard10}
{Richard} J.,  {Smith} G.~P.,  {Kneib} J.-P.,  {Ellis} R.~S.,  {Sanderson}
  A.~J.~R.,  {Pei} L.,  {Targett} T.~A.,  {Sand} D.~J.,  {Swinbank} A.~M.,
  {Dannerbauer} H.,  {Mazzotta} P.,  {Limousin} M.,  {Egami} E.,  {Jullo} E.,
  {Hamilton-Morris} V.,    {Moran} S.~M.,  2010, \mnras, 404, 325

\bibitem[\protect\citeauthoryear{{Sand}, {Treu}, {Smith} \& {Ellis}}{{Sand}
  et~al.}{2004}]{sand04}
{Sand} D.~J.,  {Treu} T.,  {Smith} G.~P.,    {Ellis} R.~S.,  2004, \apj, 604,
  88

\bibitem[\protect\citeauthoryear{{Shaw}, {Weller}, {Ostriker} \& {Bode}}{{Shaw}
  et~al.}{2007}]{shaw07}
{Shaw} L.~D.,  {Weller} J.,  {Ostriker} J.~P.,    {Bode} P.,  2007, \apj, 659,
  1082

\bibitem[\protect\citeauthoryear{{Sheth} \& {Tormen}}{{Sheth} \&
  {Tormen}}{1999}]{sheth99b}
{Sheth} R.~K.,  {Tormen} G.,  1999, \mnras, 308, 119

\bibitem[\protect\citeauthoryear{{Sheth} \& {Tormen}}{{Sheth} \&
  {Tormen}}{2002}]{sheth02}
{Sheth} R.~K.,  {Tormen} G.,  2002, \mnras, 329, 61

\bibitem[\protect\citeauthoryear{{Sheth} \& {Tormen}}{{Sheth} \&
  {Tormen}}{2004a}]{sheth04a}
{Sheth} R.~K.,  {Tormen} G.,  2004a, \mnras, 349, 1464

\bibitem[\protect\citeauthoryear{{Sheth} \& {Tormen}}{{Sheth} \&
  {Tormen}}{2004b}]{sheth04b}
{Sheth} R.~K.,  {Tormen} G.,  2004b, \mnras, 350, 1385

\bibitem[\protect\citeauthoryear{{Smith} \& {Taylor}}{{Smith} \&
  {Taylor}}{2008}]{smith08}
{Smith} G.~P.,  {Taylor} J.~E.,  2008, \apjl, 682, L73

\bibitem[\protect\citeauthoryear{{Springel}, {White}, {Tormen} \&
  {Kauffmann}}{{Springel} et~al.}{2001b}]{springel01b}
{Springel} V.,  {White} S.~D.~M.,  {Tormen} G.,    {Kauffmann} G.,  2001b,
  \mnras, 328, 726

\bibitem[\protect\citeauthoryear{{Tormen}, {Diaferio} \& {Syer}}{{Tormen}
  et~al.}{1998}]{tormen98b}
{Tormen} G.,  {Diaferio} A.,    {Syer} D.,  1998, \mnras, 299, 728

\bibitem[\protect\citeauthoryear{{Torri}, {Meneghetti}, {Bartelmann},
  {Moscardini}, {Rasia} \& {Tormen}}{{Torri} et~al.}{2004}]{torri04}
{Torri} E.,  {Meneghetti} M.,  {Bartelmann} M.,  {Moscardini} L.,  {Rasia} E.,
    {Tormen} G.,  2004, \mnras, 349, 476

\bibitem[\protect\citeauthoryear{{Umetsu}, {Broadhurst}, {Zitrin},
  {Medezinski}, {Coe} \& {Postman}}{{Umetsu} et~al.}{2011}]{umetsu11}
{Umetsu} K.,  {Broadhurst} T.,  {Zitrin} A.,  {Medezinski} E.,  {Coe} D.,
  {Postman} M.,  2011, ArXiv e-prints

\bibitem[\protect\citeauthoryear{{van de Ven}, {Mandelbaum} \& {Keeton}}{{van
  de Ven} et~al.}{2009}]{vandeven09}
{van de Ven} G.,  {Mandelbaum} R.,    {Keeton} C.~R.,  2009, \mnras, 398, 607

\bibitem[\protect\citeauthoryear{{van den Bosch}, {Norberg}, {Mo} \&
  {Yang}}{{van den Bosch} et~al.}{2004}]{vandenbosch04}
{van den Bosch} F.~C.,  {Norberg} P.,  {Mo} H.~J.,    {Yang} X.,  2004, \mnras,
  352, 1302

\bibitem[\protect\citeauthoryear{{van den Bosch}, {Tormen} \& {Giocoli}}{{van
  den Bosch} et~al.}{2005}]{vandenbosch05}
{van den Bosch} F.~C.,  {Tormen} G.,    {Giocoli} C.,  2005, \mnras, 359, 1029

\bibitem[\protect\citeauthoryear{{Wang}, {Li}, {Kauffmann} \& {De
  Lucia}}{{Wang} et~al.}{2006}]{wang06}
{Wang} L.,  {Li} C.,  {Kauffmann} G.,    {De Lucia} G.,  2006, \mnras, 371, 537

\bibitem[\protect\citeauthoryear{{Wechsler}, {Bullock}, {Primack}, {Kravtsov}
  \& {Dekel}}{{Wechsler} et~al.}{2002}]{wechsler02}
{Wechsler} R.~H.,  {Bullock} J.~S.,  {Primack} J.~R.,  {Kravtsov} A.~V.,
  {Dekel} A.,  2002, \apj, 568, 52

\bibitem[\protect\citeauthoryear{{Wu}, {Fang} \& {Xu}}{{Wu}
  et~al.}{1998}]{wu98}
{Wu} X.-P.,  {Fang} L.-Z.,    {Xu} W.,  1998, \aap, 338, 813

\bibitem[\protect\citeauthoryear{{Zhao}, {Jing}, {Mo} \& {Bn{\"o}rner}}{{Zhao}
  et~al.}{2009}]{zhao09}
{Zhao} D.~H.,  {Jing} Y.~P.,  {Mo} H.~J.,    {Bn{\"o}rner} G.,  2009, \apj,
  707, 354

\bibitem[\protect\citeauthoryear{{Zhao}, {Jing}, {Mo} \& {B{\"o}rner}}{{Zhao}
  et~al.}{2003b}]{zhao03b}
{Zhao} D.~H.,  {Jing} Y.~P.,  {Mo} H.~J.,    {B{\"o}rner} G.,  2003b, \apjl,
  597, L9

\bibitem[\protect\citeauthoryear{{Zhao}, {Mo}, {Jing} \& {B{\"o}rner}}{{Zhao}
  et~al.}{2003a}]{zhao03a}
{Zhao} D.~H.,  {Mo} H.~J.,  {Jing} Y.~P.,    {B{\"o}rner} G.,  2003a, \mnras,
  339, 12

\bibitem[\protect\citeauthoryear{{Zitrin}, {Broadhurst}, {Barkana}, {Rephaeli}
  \& {Ben{\'{\i}}tez}}{{Zitrin} et~al.}{2011a}]{zitrin11a}
{Zitrin} A.,  {Broadhurst} T.,  {Barkana} R.,  {Rephaeli} Y.,
  {Ben{\'{\i}}tez} N.,  2011a, \mnras, 410, 1939

\bibitem[\protect\citeauthoryear{{Zitrin}, {Broadhurst}, {Bartelmann},
  {Rephaeli}, {Oguri}, {Ben{\'{\i}}tez}, {Hao} \& {Umetsu}}{{Zitrin}
  et~al.}{2011c}]{zitrin11c}
{Zitrin} A.,  {Broadhurst} T.,  {Bartelmann} M.,  {Rephaeli} Y.,  {Oguri} M.,
  {Ben{\'{\i}}tez} N.,  {Hao} J.,    {Umetsu} K.,  2011c, ArXiv e-prints

\end{thebibliography}
%\bibliography{cgiocoli}
\label{lastpage}
\end{document}